\documentclass[twocolumn]{aastex631}

\usepackage{graphicx}
\usepackage{subfigure}
\usepackage{amsmath}
\usepackage{appendix}

\usepackage[mathscr]{euscript}
\usepackage{xspace}

\newcommand{\Planck}{\textit{Planck}\xspace}
\newcommand{\WMAP}{\textit{WMAP}\xspace}

\newcommand{\ymap}{$y$-map\xspace}
\newcommand{\ymaps}{$y$-maps\xspace}

\defcitealias{KTully17}{KT17}
\defcitealias{Arnaud10}{A10}
\defcitealias{LeBrun15}{L15}

\bibpunct[; ]{(}{)}{,}{a}{}{;}

\graphicspath{{./}{figures/}}

\shorttitle{Deep-NILC}
\shortauthors{Pratt et al.}

\begin{document}

\title{Optimizing NILC Extractions of the Thermal Sunyaev-Zel'dovich Effect with Deep Learning}

\author{Cameron T. Pratt}
\author{Zhijie Qu}
\author{Joel N. Bregman}
\author{Christopher J. Miller}
\affil{Department of Astronomy, University of Michigan, Ann Arbor, MI 48104}

\begin{abstract}

All-sky maps of the thermal Sunyaev\textendash Zel'dovich effect (SZ) tend to suffer from systematic features arising from the component separation techniques used to extract the signal. In this work, we investigate one of these methods known as {\it{needlet internal linear combination}} (NILC) and test its performance on simulated data. We show that NILC estimates are strongly affected by the choice of the spatial localization parameter ($\Gamma$), which controls a bias-variance trade-off. Typically, NILC extractions assume a fixed value of $\Gamma$ over the entire sky, but we show there exists an optimal $\Gamma$ that depends on the SZ signal strength and local contamination properties. Then we calculate the NILC solutions for multiple values of $\Gamma$ and feed the results into a neural network to predict the SZ signal. This extraction method, which we call {\it{Deep-NILC}}, is tested against a set of validation data, including recovered radial profiles of resolved systems. Our main result is that Deep-NILC offers significant improvements over choosing fixed values of $\Gamma$.  

\end{abstract}

\keywords{Sunyaev\textendash Zel'dovich effect --- Component Separation --- Deep Learning}

\section{Introduction}

The hot and diffuse gas that pervades the Universe is most easily observed either from X-ray emission or the thermal Sunyaev\textendash Zel'dovich (SZ) effect. The SZ effect is caused by hot electrons giving an energy boost to cosmic microwave background (CMB) photons through inverse Compton scattering. It has a weaker dependence on the gas density (linear vs. quadratic) compared to X-rays, making it an excellent probe of more diffuse gas.

The \Planck mission has provided the best data to create all-sky maps of the SZ effect (known as \ymaps). In 2016, \citet{PlanckYMAP} released the first publicly available \ymaps, nearly a half of a century after it was first theorized \citep{SZ1970,SZ1972}. Extracting the SZ signal is no easy feat as the microwave sky is dominated by non-SZ components. Over the past few decades, various component separation techniques have been developed to separate these from the SZ signal. 

These public \ymaps were constructed using {\it{internal linear combination}} (ILC) methods \citep{DelabrouilleCardoso07,Leach08}. An ILC method is a semi-blind extraction that relies on the spectral dependence of the SZ signal and uses second-order statistics to minimize the reconstruction error \citep{Delabrouille09}. ILCs also rely on a strong assumption that all non-SZ components are uncorrelated with the SZ effect. Such assumptions are not always valid and, in turn, could lead one to obtain unreliable results. If the non-SZ components are poorly separated, they manifest as systematic residuals that contaminate the reconstructed SZ signal. 

In light of this, modified ILC algorithms were developed. Modified ILC methods help improve SZ extractions by using data localization. Data localization introduces additional assumptions about the non-SZ signals before applying ILC. Among the most popular is the algorithm known as {\it{needlet internal linear comibination}} (NILC). NILC uses harmonic and spatial localization. It assumes non-SZ components are more similar when grouped by their angular size and location on the sky.  

There also exists many other component separation techniques that rely on their own assumptions (many are listed at this website\footnote{https://lambda.gsfc.nasa.gov/toolbox/comp\_separation.html}). For example, \citet{Bobin08} used the assumption of sparsity to perform a semi-blind extraction, known as general morpholocial component analysis (GMCA), of the CMB and SZ signals. While GMCA could outperform a few other extraction techniques, it yielded similar results to NILC. Over the years, refinements have been made to improve its performance \citep{Bobin15,HGMCA}. One could also do parametric modeling of the components, which has largely been done with the {\it{Commander}} algorithm \citep{Eriksen08,Galloway23}. While many more component separation techniques exist, most have only been applied to the CMB. Only recently have there been efforts to improve the quality of all-sky SZ data \citep[e.g.,][]{McCarthy23}, so there is still plenty of room for investigations.

One modern approach to improving SZ extractions is with deep learning (DL). DL is a subset of machine learning techniques that uses neural networks to build non-linear models. They excel at capturing complex relations between variables in a way that simple, traditional models cannot describe. Over the last decade, DL has been used to tackle challenging tasks, and has proven to perform well in various facets of astrophysics including: galaxy morphology classification \citep{Zhu19}, detection of exoplanets \citep{Shallue18}, painting baryons onto dark matter halos in cosmological simulations \citep{Troster19}, separating the CMB from foreground contamination \citep{Casas22}, and building catalogs of SZ clusters \citep{Voskresenskaia23}. To our knowledge, DL has not been used to extract the SZ effect, and this is one of the chief goals of this paper.

In this work, we present the NILC method and specifically focus on its spatial localization parameter, $\Gamma$. Traditional implementations of NILC fix the value of $\Gamma$, however, we demonstrate that yields a sub-optimal solution in certain scenarios. Then we present a novel method for extracting the SZ signal which we call {\it{Deep-NILC}}. Deep-NILC combines NILC results from different values of $\Gamma$, then feeds these values into a neural network to predict the true SZ signal. Our results show that we can achieve better estimations of the SZ signal compared to using fixed values of $\Gamma$. 
 
The paper is structured in the following way: in \autoref{sec:methods} we present the methodology for this study; the main results are presented in \autoref{sec:results}; the results are discussed in \autoref{sec:discussion}; a summary is provided in \autoref{sec:summary}.

\section{Methods} \label{sec:methods}
\subsection{Simulations} \label{sec:simulatoins}
The main data used in this study consisted of simulated observations of the microwave sky. First we describe the set of simulated \ymaps obtained from \citet{Han21} used for training and validation. Then we inject additional SZ sources to better represent the nearby ``anomalies'' that exist in the real Universe. Finally, we generate realistic mock observations as seen by \Planck and \WMAP.

\subsubsection{SZ Data from Han et al. (2021)}
This work required a sizable set of simulated \ymaps that resembled the statistical properties of our observed Universe. Usually, a simulated \ymap would be acquired from full 3-D cosmological simulations by projecting the electron pressure along a line-of-sight. Unfortunately, running a single simulation can be very computationally expensive, so there are only a few available for public use.

We were able to overcome this obstacle using the work of \citet{Han21}. These authors generated a large set of mock extragalactic signals at millimeter wavelengths based off the cosmological simulations of \citet{Sehgal10}. The simulations from \citet{Sehgal10} were designed to match the best observational data available at the time, using cosmological parameters inferred from \WMAP and gas prescriptions that matched X-ray observations. \citet{Han21} developed an algorithm, known as {\it{MillimeterDL}} (mmDL), to generate many synthetic \ymaps using a single simulation from \citet{Sehgal10}. 

The mmDL algorithm used a DL method known as Generative Adversarial Networks (GANs). GANs consists of two components: a generator and a discriminator. The generator generates synthetic data while the discriminator's role is to distinguish between real and fake data. Through an iterative process, the generator improves its ability to produce realistic data by fooling the discriminator until the mock data are nearly indistinguishable from the input training data. mmDL was designed to match the power spectra from the \citet{Sehgal10} simulations, but it was also able to capture important non-Gaussian features. More generally, mmDL was built to take any cosmological simulation as input and generate a large set of corresponding realizations. By feeding a simulation into mmDL as input, \citet{Han21} created 500 realizations that are publicly available\footnote{https://portal.nersc.gov/project/cmb/data/generic/mmDL/}. 

Their SZ effect was provided at different frequencies which we converted into \ymaps using the known frequency dependence of the SZ signal (see \autoref{eq:sz_freq} and \autoref{eq:sz_temp}). Furthermore, only ten realizations of the SZ effect were used for our experiments. Seven of these were used for supervised training while three were set aside for validation. In the next section, we augment these data with more examples of resolved signals. 

\subsubsection{Resolved SZ Signals} \label{sec:resolved}
Most of the strong SZ signals in the Universe are unresolved by \Planck and \WMAP. Although, there exists a few extended objects, and these are amongst the most interesting SZ signals to study. For example, the Virgo cluster is somewhat of a statistical anomaly as it produces a strong SZ profile that extends many degrees on the sky. In addition, there are nearby low-mass clusters, galaxy groups, and individual galaxies that also produce resolved signals. Such systems have been carefully looked at in X-rays and SZ in hopes of closing the baryon budget and understanding various feedback cycles \citep{Bregman18, Pratt21, Bregman22}.

When constructing the training data, it appeared the simulations of \citet{Han21} did not contain enough examples of the unique, resolved SZ sources observed in the real Universe. In order to overcome this, we created our own simulated set of resolved signals. These signals were generated based on X-ray observations reported in the {\it{Meta-Catalog of X-Ray Detected Clusters of Galaxies}} \citep[MCXC;][]{MCXC}. We used the reported masses ($M_{500}$\footnote{The subscript ``500'' denotes the property at which the density equals 500 times that of the critical density.}) and redshifts ($z$) and then calculated the angular size of $R_{500}$ for each system. Resolved systems were selected as those with $R_{500} > 30^{'}$ which yielded a sample of 25 objects.

The next step was to simulate the SZ profiles of the resolved signals. They were generated using the universal pressure profile ($P(r)$) from the AGN 8.0 simulations of \citet{LeBrun15}. Their SZ surface brightness profiles, $y(r)$, as a function of physical radius is given by
\begin{equation}
    y(r) = \frac{\sigma_{\mathrm{T}}}{m_{\mathrm{e}}c^{2}} \int_{r}^{R_{max}} \frac{2P(r')r'}{\sqrt{r^{'2} - r^{2}}}dr'
\end{equation}
where $R_{max} = 5 R_{500}$, $\sigma_{\mathrm{T}}$ is the Thompson cross-section, $m_{\mathrm{e}}$ is the electron rest-mass, and $c$ is the speed of light. These profiles were then superimposed with the simulated \ymaps from \citet{Han21}. Herein, we denote the resolved part of the simulated \ymaps as the resolved SZ component and that from \citet{Han21} as background SZ. There were still a few resolved signals in \citet{Han21}, but we use this demarcation to distinguish the two components.

Then we superimposed the resolved profiles with the SZ background. Before injecting the signals, we added some randomization to help with generalization. Each object was placed at different distances by randomly simulating their signals at redshifts of $\pm 50\%$ their MCXC values. Furthermore, these were injected into the northern and southern Galactic hemispheres at random positions of moderately high Galactic latitudes ($|b|>20^{\circ}$). This was to simply augment the sample while also avoiding the highly contaminating regions near the Galactic plane. 

The resolved signals used for validation consisted of two data sets. The first was constructed in exactly the same way as the training data but using the three background SZ maps that were originally set aside. These were used for statistical analyses in \autoref{sec:results}. The second set of validation data were made in order to inspect specific signals: a nearby galaxy group of $M_{500}=10^{13} M_{\odot}$ at 10 Mpc (R$_{500} = 113^{\prime}$), a signal resembling the Coma cluster (R$_{500} = 47^{\prime}$), and that of the Virgo cluster (R$_{500} = 177^{\prime}$). These SZ profiles were also created using the AGN 8.0 pressure profile. 

For the Coma- and Virgo-like sytems, ten realizations were generated with five injected both into the northern and southern Galactic hemispheres. For simplicity, we placed them at Galactic latitudes of $b = \pm 50^{\circ}$ and spaced them evenly in Galactic longitude starting at $l = 0^{\circ}$. For the galaxy group-like signal, we generated fifty signals and randomly placed them across the sky with a Galactic latitude restriction of $|b|>20^{\circ}$. Separate \ymaps were generated for each of the three systems, but they were superimposed with the same background SZ for consistency. These validation data were used to investigate the recovered radial profiles in \autoref{sec:results}.

\subsubsection{Planck Sky Model} \label{sec:pysm}
Here we describe the modeling of non-SZ emissions, which we call contaminants. These were generated using the Planck Sky Model \citep{PlanckSkyModel} version that has been wrapped into Python \citep[PYSM;][]{Thorne2017}. The Planck Sky Model was developed for simulating microwave radiation observed by the Planck satellite. This software was mainly developed to help researchers test their pipelines and validate cosmological models against observational data. 

The components \footnote{https://pysm3.readthedocs.io/en/latest/index.html} used in this work included: Galactic thermal dust emission, synchrotron emission, anomalous microwave emission, free-free emission, the lensed CMB, CO line emission, cosmic infrared background (CIB), and radio galaxies. For some of these components, there was only one available model in PySM which included the SZ effect. A single realization of the SZ effect was not sufficient for our purposes which is why we incorporated the data products from \citet{Han21}. A consistent foreground model (d1-s1-a1-f1-c2-co1-cib1-rg1 components from PYSM) was used for the non-SZ components in all ten realizations. Therefore, the only differences between the simulated all-sky maps were the distributions of the SZ signal. The components were then integrated over the \Planck and \WMAP frequency response functions to simulate realistic observations. We implemented the publicly available transmission curves both for \Planck \footnote{https://portal.nersc.gov/cfs/cmb/planck2020/misc/} and \WMAP \footnote{https://lambda.gsfc.nasa.gov/product/wmap/dr5/bandpass\_get.html} at their respective frequencies. Lastly, all of the frequency maps were produced in {\it{HEALPix}} format \citep{Healpix} with a resolution parameter $\mathrm{NSIDE}=1410$ which yields four pixels per resolution element of $10^{\prime}$. Instrumental noise and beam effects were ignored for simplicity, but they may be worth including in future work.

\subsection{SZ Signal Extraction}
In this section, we explain the methods used to extract the SZ signal. We first present NILC in detail. Most importantly, we demonstrate the significance of its spatial parameter, $\Gamma$, that controls a bias-variance trade-off. Then we use the data products from multiple NILC extractions as input to a neural network called Deep-NILC. In the following, bold symbols indicate matrices or vectors.

\subsubsection{Internal Linear Combination (ILC)}\label{sec:ILC}
The SZ effect boosts the energy of low-energy CMB photons ($\nu < 217$ GHz) to higher energies. This causes a distortion in the CMB spectrum that is well-grounded in theory and is given by
\begin{equation}
\label{eq:sz_freq}
    g_{\nu} = g(x) = \bigg( x \frac{e^{x}+1}{e^{x}-1} - 4\bigg)
\end{equation}
where $x = \frac{h\nu}{k_{\mathrm{B}}T_{\mathrm{CMB}}}$ ($k_{\mathrm{B}}$ is the Boltzmann constant, $h$ is Planck's constant, and $T_{\mathrm{CMB}}$ is the temperature of the CMB). The amplitude of the SZ effect is measured as a dimensionless quantity known as the Compton parameter, $y$, given by
\begin{equation}\label{eq:sz_temp}
    \frac{\Delta T}{T_{\mathrm{CMB}}} = g_\nu y
\end{equation}
where $g(\nu)$ is the SZ spectral dependence, $\frac{\Delta T}{T_{\mathrm{CMB}}}$ is the change in temperature relative to the CMB.

In practice, the SZ signal is observed alongside other cosmological signals and Galactic emissions in the radio/IR bands. The observed intensity, $X_{\nu}$, for each frequency channel, $\nu$, at pixel, $p$, can be written as a sum of various components
\begin{equation}
\label{eq:ILC}
    X_{\nu}(p) = g_{\nu}y(p) + N_{\nu}(p)
\end{equation}
where $N_{\nu}(p)$ denotes the signals from all non-SZ components, including instrumental noise as well as contaminating astrophysical signals. One can also rewrite \autoref{eq:ILC} as
\begin{equation}
\label{eq:ILC_delta}
    \Delta X_{\nu}(p) = \Delta g_{\nu}y(p) +\Delta N_{\nu}(p)
\end{equation}
where $\Delta$ represents the difference between two frequency bands. When $X_{\nu}$ is expressed in thermodynamic temperature units, the CMB has the same amplitude in each band and is deprojected from $\Delta N$. This strategy was used by \citet{Pratt21} where these authors subtracted subsequent frequency maps from the same instruments, and we applied their same method in this work. In the following equations, however, we drop $\Delta$ in our notation since \autoref{eq:ILC} and \autoref{eq:ILC_delta} are mathematically similar. 

ILC groups all non-SZ signals into a single noise term such that 
\begin{equation}
\label{eq:yhat_linear}
\vspace{-0.1cm}
    \widehat{y(p)} = y(p) +\sum_{\nu}w_{\nu}(p)N_{\nu}(p)
\vspace{-0.1cm}
\end{equation}
where $\widehat{y}$ is the estimated SZ signal, and $w_{\nu}$ are the weights that minimize the total covariance projected along the $g(\nu)$ direction while also preserving the signal i.e., $\sum_{\nu} g_{\nu}w_{\nu} = \mathbf{g} \mathbf{w}^{T} = 1$. These weights can be solved for with Lagrange multipliers
\begin{equation}\label{eq:ilc_weights}
    \mathbf{w} = \frac{\mathbf{g^{T} R^{-1}}}{\mathbf{g^{T} R^{-1} g}}
\end{equation}
and the estimator of the SZ signal is
\begin{equation}\label{eq:ilc_estimate}
    \widehat{y} = \mathbf{w^{T}} \mathbf{X}
\end{equation}
where \textbf{X} is the matrix of frequency observations, \textbf{R} is the covariance matrix \textbf{R} = \textbf{X}\textbf{X}$^{T}$, and \textbf{w} is the vector of calculated weights. 

The ILC solution completely depends on the construction of $\mathbf{R}$, which is derived from the data themselves. In the standard (simplest) ILC method, one uses all available data over the entire sky to calculate the weights. This assumes the spectral properties of the non-SZ emissions do not change across the sky. Such an assumption is naive since the contaminants are known to be non-stationary. In turn, the standard ILC method poorly separates the contaminants from the SZ signal. 

One can achieve better ILC extractions by localizing data when constructing $\mathbf{R}$. Localization refers to the sub-grouping of data, under certain assumptions, to calculate the ILC weights. When the localization is strong\textemdash that is when data are finely segmented\textemdash the propagation of contamination can be suppressed. This reduces the variance in the reconstructed \ymap. Conversely, strong localization renders a biased result when the localized contamination and SZ signal are correlated. While many of the contaminating signals may not be physically correlated with the SZ effect, random empirical correlations still exist. 

This bias has been discussed in previous studies \citep{Delabrouille09, Remazeilles13, McCarthy23} and we present an overview for completeness. The estimated SZ signal can be written as
\begin{equation}
\label{eq:yhat_var}
    \langle \widehat{y}^{2} \rangle = \langle (y + \widehat{N})^2 \rangle = \langle y^{2} \rangle + 2\langle y\widehat{N}\rangle + \langle \widehat{N}^{2} \rangle
\end{equation}
where $\widehat{N} = \sum_{\nu}w_{\nu}N_{\nu}$ equals the last term in \autoref{eq:yhat_linear} and denotes the total amount of reconstructed noise. There is a correlation term between the SZ signal and the noise known as the ILC bias defined as
\begin{equation}
    \langle y\widehat{N} \rangle = \frac{-\langle y^{2}\rangle(m-1)}{N_{p}}
    \label{eq:ILC_bias}
\end{equation}
where $m$ denotes the number of available frequency maps (or difference maps in this study) and $N_{p}$ is the number of modes used to construct $\mathbf{R}$ \citep[see derivation by][]{Delabrouille09}. Note that this term is negative and, thus, reduces the power of the SZ signal. Rearranging \autoref{eq:yhat_var} we get
\begin{equation}
    \langle \widehat{y}^{2} \rangle = \langle y^{2} \rangle - \frac{2\langle y^{2}\rangle(m-1)}{N_{p}} + \langle\widehat{N}^{2}\rangle.
\end{equation}
An optimal solution is reached when the sum of the last two terms is zero, or equivalently when
\begin{equation}
\label{eq:bias_var_tradeoff}
    2(m-1)\langle y^{2}\rangle = {N_{p}}\langle \widehat{N}^{2}\rangle.
\end{equation}

It is important to note that the reconstruction noise depends on $N_{p}$. Empirically, $\langle \widehat{N}^{2}\rangle$ and $N_{p}$ are positively correlated. Their relation, however, cannot be easily described because $\langle \widehat{N}^{2}\rangle$ depends on the changing the contamination properties across the sky. This positive correlation suggests when $N_{p}\langle \widehat{N}^{2}\rangle$ is large, an optimal balance can be achieved when $\langle y^{2} \rangle$ is also large. This is a key point: one can afford a larger reconstruction error if the underlying signal is strong enough. On the other hand, if the true signal is weak, then one should consider reducing $N_{p}$ to achieve a better solution. 

\begin{figure*}
     
    \centering
    \includegraphics[width=0.95\textwidth]{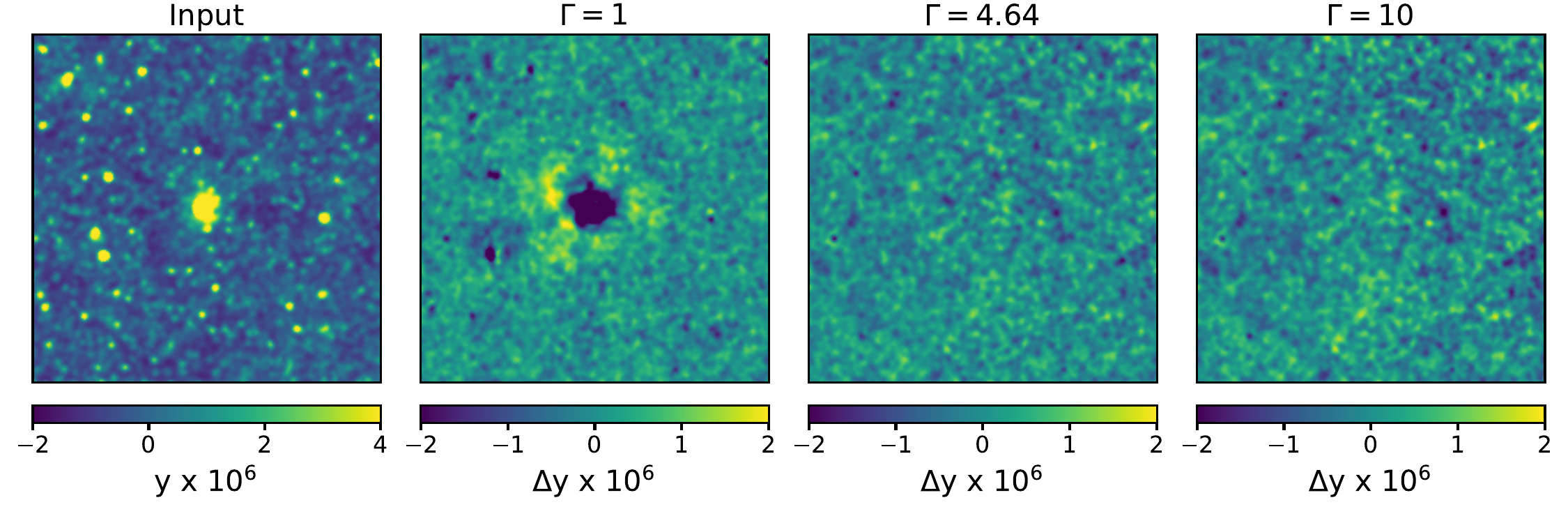}
    \centering
    \includegraphics[width=0.98\textwidth]{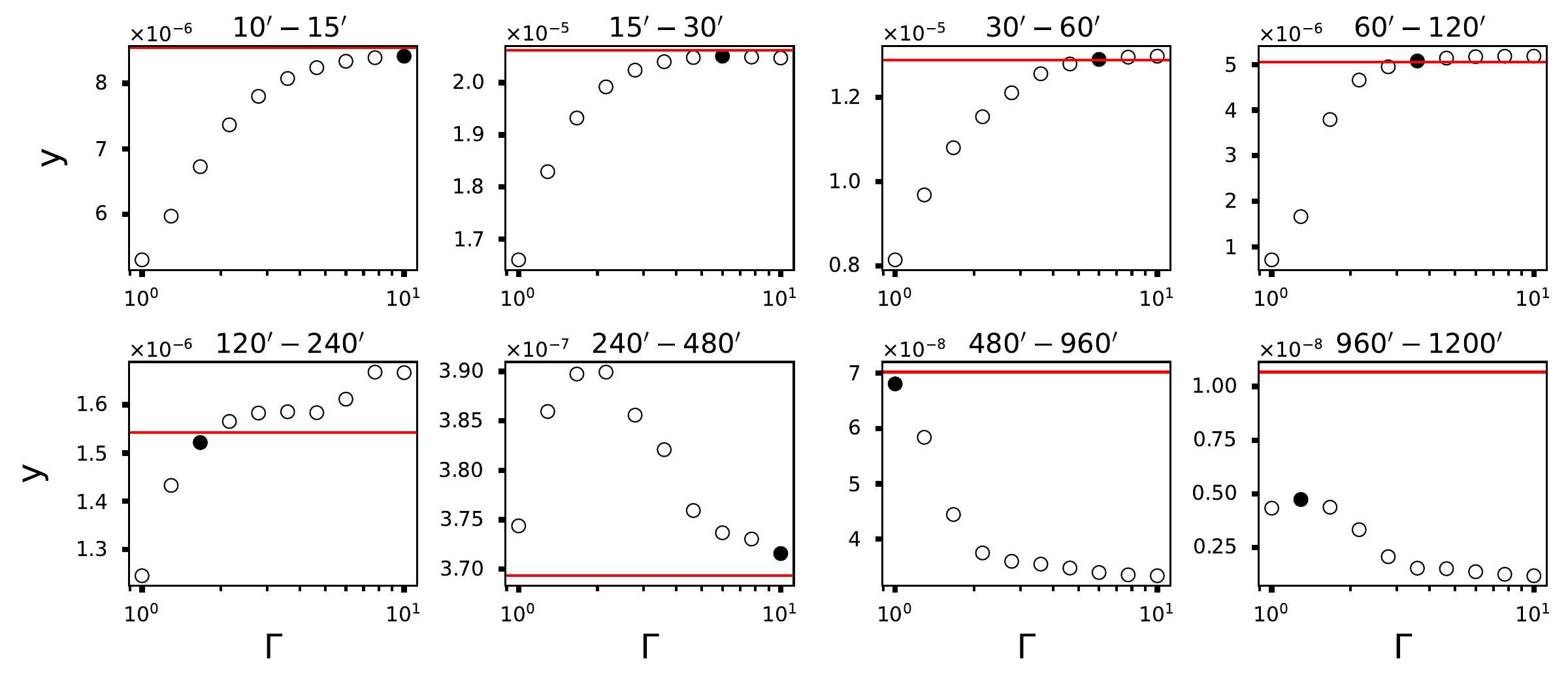}
    
    \caption{An example of how different values of $\Gamma$ affect the NILC extraction of the SZ signal. The images in the top row are $10^{\circ}x10^{\circ}$ and centered around a strong simulated SZ source. The leftmost panel denotes the input SZ signal. The panels to the right show residuals between the input and reconstructed \ymaps using three values of $\Gamma$. The eight panels below display the different estimates of the center pixel in the above images. They are shown as a function of wavelet scale and $\Gamma$. The best possible NILC solution given an array of $\Gamma$ values are shown as filled circles; this is the closest estimate to the true value which is shown as a horizontal line in red. Summing the true values (red lines) on each wavelet scale yields the total SZ signal. The main point of this plot is to demonstrate how the choice of $\Gamma$ impacts the estimate of the SZ signal. It also shows that larger values of $\Gamma$ are typically preferred when performing extractions where the SZ signal is strong.}
    
\label{fig:example_cluster}
\end{figure*}


\begin{figure*}
     
    \centering
    \includegraphics[width=0.95\textwidth]{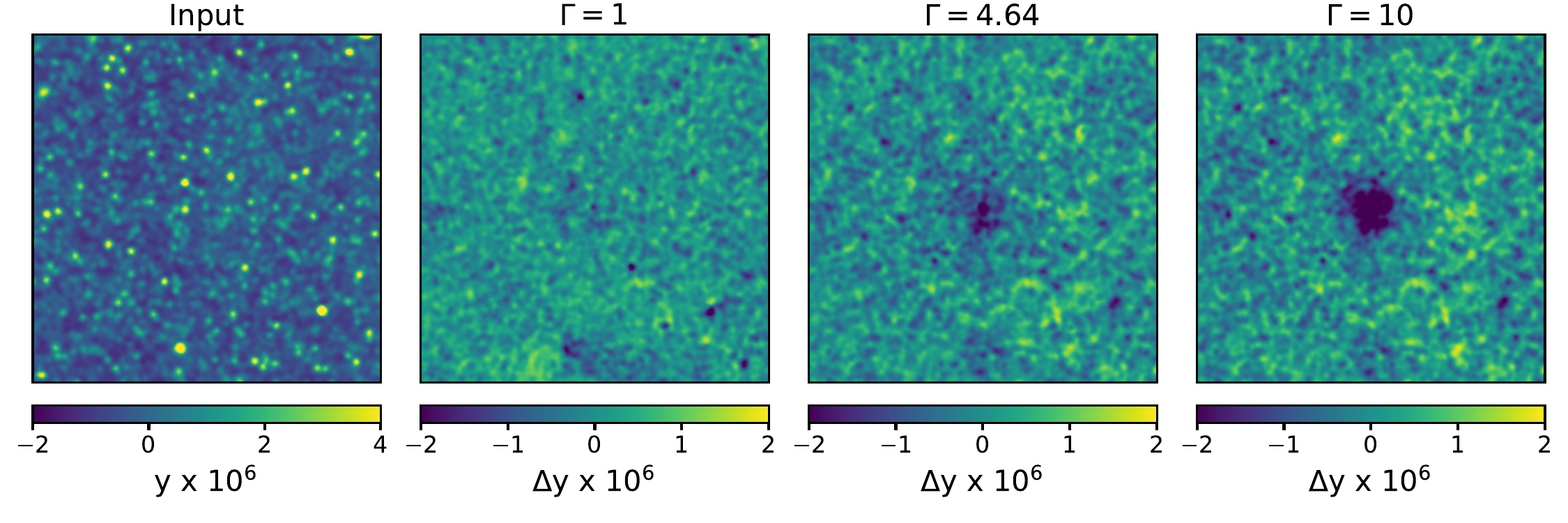}
    \centering
    \includegraphics[width=0.98\textwidth]{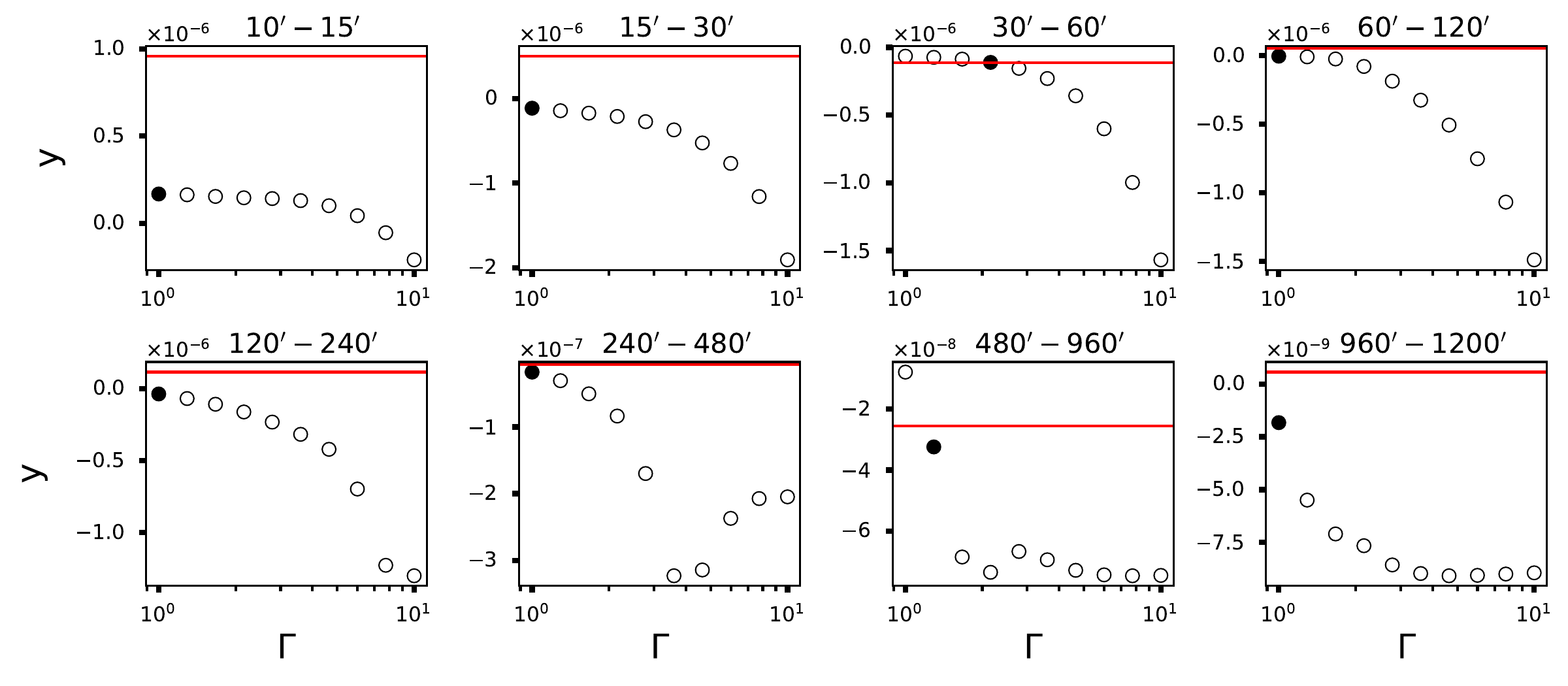}
    
    \caption{Same as \autoref{fig:example_cluster} except for the change in location of the sky. This example is centered around a pixel with a weak SZ signal, but it is also centered around a known radio source from the simulations. Smaller values of $\Gamma$ were preferred in this scenario, which were generally true for pixels with weak SZ signals and/or strongly contaminated regions.}
    
\label{fig:example_radio}
\end{figure*}

\subsubsection{Needlet Internal Linear Combination (NILC)}\label{sec:NILC}

Among the most popular localization methods is the NILC algorithm, a modified ILC algorithm that controls $N_{p}$ i.e., the strength of data localization. NILC harmonically and spatially localizes data on the sphere \citep{Guilloux07} by clustering them based on angular size and location on the sky. In practice, one first transforms the data into a wavelet basis, and then applies a spatial weighting scheme when constructing the covariance matrix in \autoref{eq:ilc_weights}. Below we describe the NILC method used throughout this work. We refer to \autoref{eq:nilc_flow} as a guide for the workflow
\begin{equation}
\label{eq:nilc_flow}
    \mathbf{X} \Longrightarrow \mathbf{\tilde{X}} \longrightarrow \mathbf{\tilde{R}} \longrightarrow \widehat{\mathbf{\tilde{y}}} \Longrightarrow \widehat{y}.
\end{equation}
The double arrows ($\Longrightarrow$) denote a transforms to/from the wavelet basis. Single arrows ($\longrightarrow$) represent processes performed on individual wavelet scales. We denote some variable, $Y$, transformed in the wavelet domain as $\Tilde{Y} = \{ \Tilde{Y}_{1},\Tilde{Y}_{2},...,\Tilde{Y}_{K} \}$, where $K$ is the number of window functions. 

The first step ($\mathbf{X} \Longrightarrow \mathbf{\tilde{X}}$) was to transform the frequency data into a wavelet basis using harmonic window functions; this is the same as \textit{Analysis} step from \citet{Delabrouille09} These consisted of subtracted Gaussian beams characterized by their full-width at half-maximum (FWHM $=\mathscr{F}$). The wavelet bands were defined as: 10$^{\prime}$-15$^{\prime}$, 15$^{\prime}$-30$^{\prime}$, 30$^{\prime}$-60$^{\prime}$, $\mathscr{F}^{u}_{i-1}$ - $2\mathscr{F}^{u}_{i-1}$, ..., 960$^{\prime}$-1200$^{\prime}$. The window functions scaled dyadically using the largest FWHM of the previous window which we define as $\mathscr{F}^{u}_{i-1}$ where $i$ indexes the wavelet bands and the superscript $u(l)$ denotes upper(lower) FWHM for each window. For clarity, we note that $\mathscr{F}^{l}_{i} = $$2\mathscr{F}^{l}_{i-1}$. However, dyadic scaling was not applied to the first and last windows because we set a maximum resolution of 10$^{\prime}$ and the largest scale at $20^{\circ}$. Typically, the standard wavelet functions, $\psi_{i}$, are defined such that $\sum_{i}^{K} |\psi_{i}|^{2} = 1$. The window functions used in this study, however, altered the sum to $\sum_{i}^{K} |\psi_{i}|^{2} = G(\mathscr{F}=10') - G(\mathscr{F}=20^{\circ})$, where $G(\mathscr{F})$ represents the transmission of the Gaussian beam.

The first half of the window functions were applied to the frequency data, transforming them into a wavelet basis i.e., $\mathbf{\Tilde{X}}_{i} = \langle \mathbf{X}, \psi_{i} \rangle$. (The second half will be applied later to satisfy the square-integrable condition.) Only some of the frequency data could be used in certain wavelet bands based on their corresponding beam sizes in either the \Planck or \WMAP missions. For example, only the data from the high-frequency instrument on \Planck could be used in the first wavelet band because the resolution was better than $10^{\prime}$. 

The next step ($\mathbf{\tilde{X}} \longrightarrow \mathbf{\tilde{R}}$) was to obtain $\tilde{\textbf{w}}$ by solving \autoref{eq:ilc_weights}. As previously mentioned, this depends entirely on the construction of $\Tilde{\textbf{R}}$. NILC spatially localizes data on a pixel-by-pixel basis by constructing $\Tilde{\textbf{R}}$ from only nearby pixels. The contribution of each pixel to $\Tilde{\textbf{R}}$ was based on an angular separation weighting scheme where closer pixels carried higher weights. We implemented a Gaussian weighting scheme ($\beta$) as a function of pixel separation, $\Delta s$, such that
    \begin{equation}
        \beta \propto e^{-\Delta s^{2}/\sigma^{2}}
    \end{equation}
where $\sigma$ controlled the amount of spatial localization. $\sigma$ was scaled with each window function such that 
\begin{equation}
    \sigma_{i} = \Gamma \times \mathscr{F}^{\mathrm{{u}}}_{i}
\end{equation}
where $\Gamma$ is the spatial localization parameter. 

The next step ($\mathbf{\tilde{R}} \longrightarrow \widehat{\mathbf{\tilde{y}}}$) solved for the SZ signal in each wavelet band using \autoref{eq:ilc_weights} and \autoref{eq:ilc_estimate}. Then one applies the second application of the wavelet function to achieve full spatial resolution. The last step ($\widehat{\mathbf{\tilde{y}}} \Longrightarrow \widehat{y}$) was the wavelet reconstruction $\widehat{y} = \sum_{i}^{K} \Tilde{y}_{i}$ which added the \ymaps per wavelet scale; this is the same as \textit{Synthesis} step from \citet{Delabrouille09}. In the next section, we skip this last step and, instead, use the \ymaps per wavelet scale as input to a neural network.

\subsubsection{Deep-NILC}

In this section we explain the pipeline to produce a \ymap with Deep-NILC. There are two main parts: the first is a data pre-processing stage using the NILC algorithm, and the second step models these data using a neural network. 

The first part of the pipeline was to apply the NILC algorithm to our set of simulated data. Unlike the traditional NILC, we did not assume some $\Gamma$ but rather performed extractions for a range of reasonable values; specifically, ten values of $\Gamma$ with logarithmic spacing from [1-10] (i.e., 1, 1.292, 1.668, ..., 10). A total of 80 decomposed \ymaps were created for a single simulation of the sky given the eight harmonic bands and ten values of $\Gamma$.

Varying $\Gamma$ provided information on how the ILC weights reacted to the surrounding contamination. One could imagine scenarios where using one value would be more suitable than others. For example, \autoref{fig:example_cluster} shows an example of how altering $\Gamma$ affected the extraction for a pixel near the center of a strong cluster. According to \autoref{eq:bias_var_tradeoff}, extracting the SZ signal around a massive cluster should be done with a large value of $\Gamma$. In this case, the signal was large enough where one becomes more concerned about the bias rather than contamination. The residual images shown in the top row of \autoref{fig:example_cluster} demonstrated that larger values of $\Gamma$ yielded better estimates near a strong SZ source. On the other hand, \autoref{fig:example_radio} shows a case where the SZ signal was weak and coincident with a known radio source from the PySM simulations. In this case, $\Gamma=1$ helped reduce the contaminating signal from the radio source. One can see the true values (red lines) did not agree well with any of the NILC solutions for the first couple of wavelets, however, the solution when $\Gamma=1$ was the closest match. Based on the arguments above, one can achieve better NILC extractions when $\Gamma$ is allowed to vary. 

Now, we seek a model that can take NILC extractions from various $\Gamma$ values to improve estimations of the SZ signal. In order to tackle this problem, we used a fully connected, feed-forward network known as a multilayer perceptron (MLP). MLPs consist of multiple layers of interconnected nodes, known as neurons. The networks are designed to process information from the input layer, through a set of hidden layers to the output layer. Each neuron receives signals from the neurons in the previous layer and returns a weighted linear combination of the inputs. The result then undergoes a non-linear activation function which is then fed as input to the next layer. The parameters of the network are then trained with supervised learning through an iterative procedure known as gradient descent. After each iteration, the trainable parameters are updated to minimize the error between the desired output and predicted output. 

The network performed regression by minimizing the mean absolute error (MAE) loss function 
\begin{equation}
    \mathrm{MAE} = |y-y^{\prime}|
\end{equation}
where $y^{\prime}$ denotes the best possible NILC solution. In other words, $y^{\prime}$ was constructed from the $\Gamma$ that yielded the closest value to the true SZ signal on each wavelet scale then summing them together. Examples of this are shown as the black circles in \autoref{fig:example_cluster} and \autoref{fig:example_radio}. We note the only purpose of $y^{'}$ was for training the MLP, and it is not available for real data. 

We chose to regress on $y^{\prime}$ instead of the true SZ values to keep our model from over-correcting, and keeping the results within the confines of the NILC method. Taking \autoref{fig:example_radio} as an example, many of the NILC predictions were far away from their true values when extracting the SZ signal in a highly contaminated region. If we regressed on the true values, this region would increase the MAE and the model would work hard to accommodate to the large residuals. Replacing the true values with $y^{\prime}$ mitigates this impact by reducing the range of residuals.

\begin{longtable}{|c|c|c|}
\caption{Neural Network Parameters} \label{tab:NN} \\
 \hline
 Input dimensions & $10$ $(\mathrm{N}_{\Gamma}) \times 8$ (K)\\
 \hline
Output dimensions & 1 \\

\hline 
Number of hidden layers & 9 \\
\hline
Number of neurons & 80, 40, 20, 20, 10, 10, 5, 5, 3\\
\hline
Activation (hidden) & Leaky ReLU\\
\hline
Activation (output) & Linear\\
\hline
Optimizer & Adam\\
\hline
Batch size & 64 \\
\hline 
Initial learning rate & 0.001\\
\hline
Loss function & MAE\\
\hline
\end{longtable}

A number of hyperparameters had to be defined, and those of our network are shown in \autoref{tab:NN}. The input layer accepted a matrix of shape (K $\times$ N$_{\Gamma}$) where $N_{\Gamma}$ represents the ten $\Gamma$ values. The data were then flattened into a vector before passing to the first hidden layer. The MLP contained nine hidden layers with Leaky Relu activation functions (slope for negative values was 0.01) after each hidden layer. The output layer used a linear activation and yielded a single number. In addition, normalization was applied to the input and output data to help train the model by scaling and shifting them to have zero mean and unit variance. We tuned our model to find optimal values of the initial learning rate, mini-batch size, number of layers, and number of nodes per layer. We decided to use the Adam optimizer \citep{Adam} with an initial learning rate of $0.001$ with exponential decay rates of 0.9 and 0.999 for the first and second moments respectively. The most sensitive hyperparameter was the learning rate. All other hyperparameters, such as the batch size and network architecture, did not require meticulous tuning. Finally, we restricted the set of training pixels to be at Galactic latitudes $|b|>10^{\circ}$ to avoid training in the most highly contaminated regions.




%
%


\section{Results}\label{sec:results}
In this section, we present the main results from the validation data. Specifically, we focus on the performance of Deep-NILC to the \ymaps generated with fixed values of $\Gamma$. We consider three nominal values of gamma to give a sense of how NILC extractions behave: $\Gamma=1, 4.64,$ and 10. (These were selected from a set of $\Gamma$ values with logarithmic scaling.) First, we show residual statistics between the estimated \ymaps and the true signal including: power spectra, bias estimates, and cross-correlations between the strongest contaminants. Then we present the radial profile extractions for resolved systems.


\begin{figure}[t!]
     
    \centering
    \includegraphics[width=0.45\textwidth]{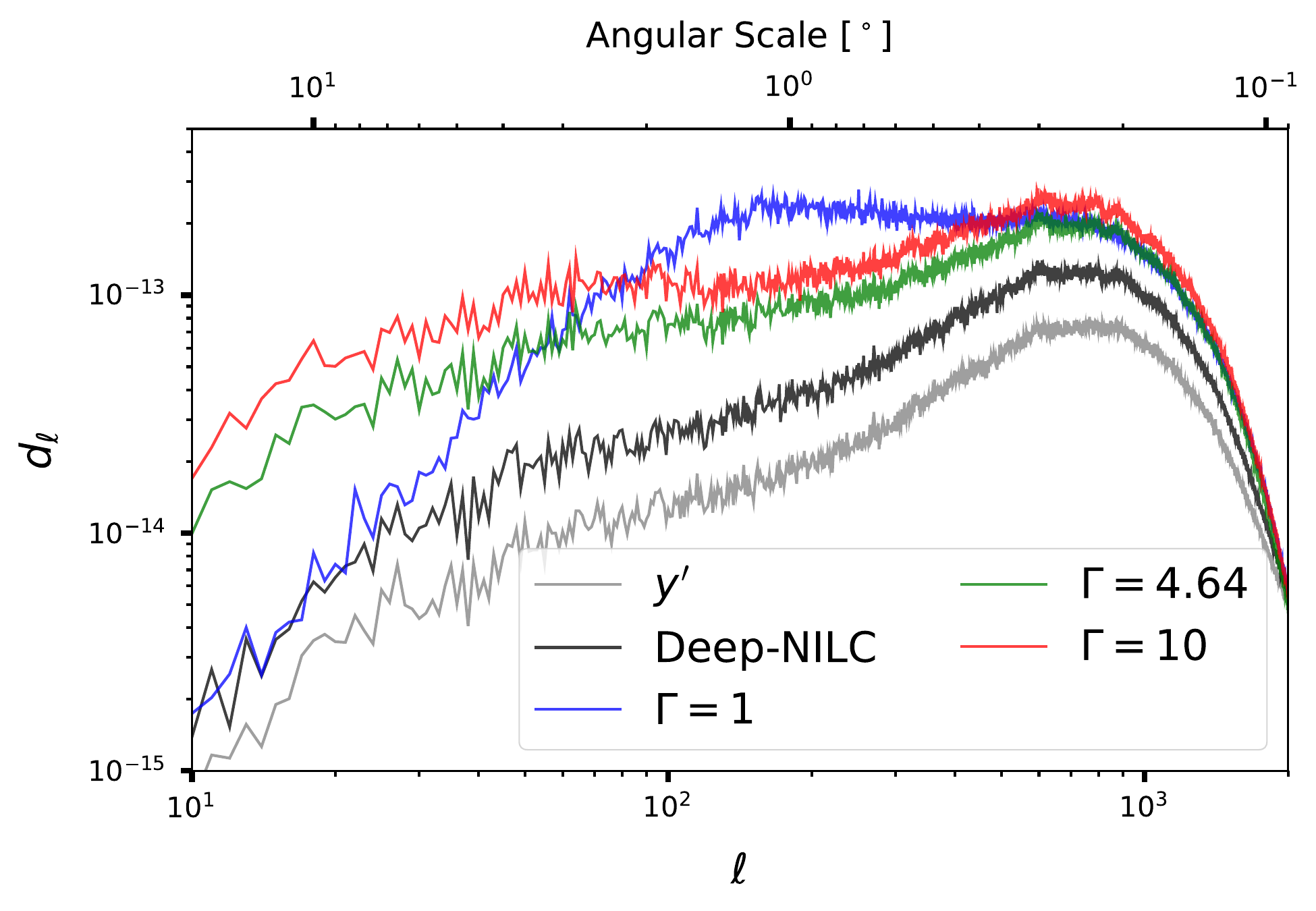}
    
    \caption{Average residual power spectra for the three validation \ymaps. The gray line denotes the residual for the best possible NILC solution and black is the results from Deep-NILC. The blue, green, and red are the residuals for $\Gamma = 1, 4.64, 10$ respectively.}
    
\label{fig:residual_power}
\end{figure}

\subsection{Residual Power Spectra}
Here we consider the power spectra of the residual maps ($\widehat{y} - y$), where $y$ is the true SZ signal. The residual power, $\mathbf{d_{\ell}}$, defined as $\frac{\ell(2\ell+1)}{2\pi}\Delta C_{\ell}$ where $\Delta C_{\ell}$ was computed from the residual map after setting pixels of low Galactic latitude ($|b|<10^{\circ}$) to zero. In \autoref{fig:residual_power}, we show the residual power for five \ymaps. The gray line shows the results for the best possible NILC solution ($y^{\prime}$); the black line denotes Deep-NILC; and the colored lines display the results for fixed values of $\Gamma$. As expected, the gray line showed the smallest amount of residual power. The blue line exhibited the largest amount of power for $\ell \sim 200$, but less power compared to the other values of $\Gamma$ when $\ell \lesssim 40$. Deep-NILC had the least amount of residual power compared to all fixed values of $\Gamma$ on all angular scales.

\subsection{Bias and Scatter}
\begin{figure*}
   \centering
   \includegraphics[width=0.48\textwidth]{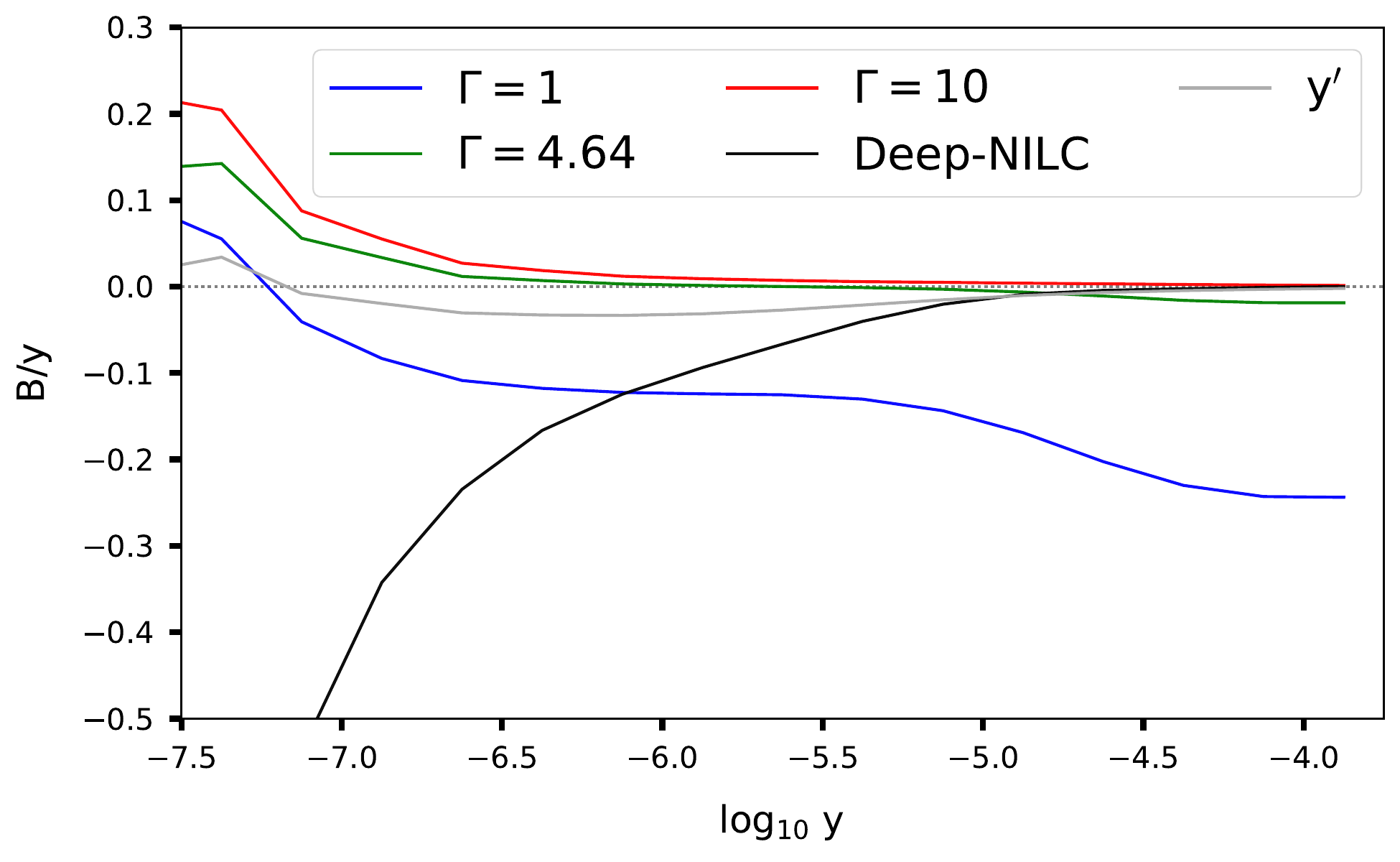}
   \includegraphics[width=0.48\textwidth]{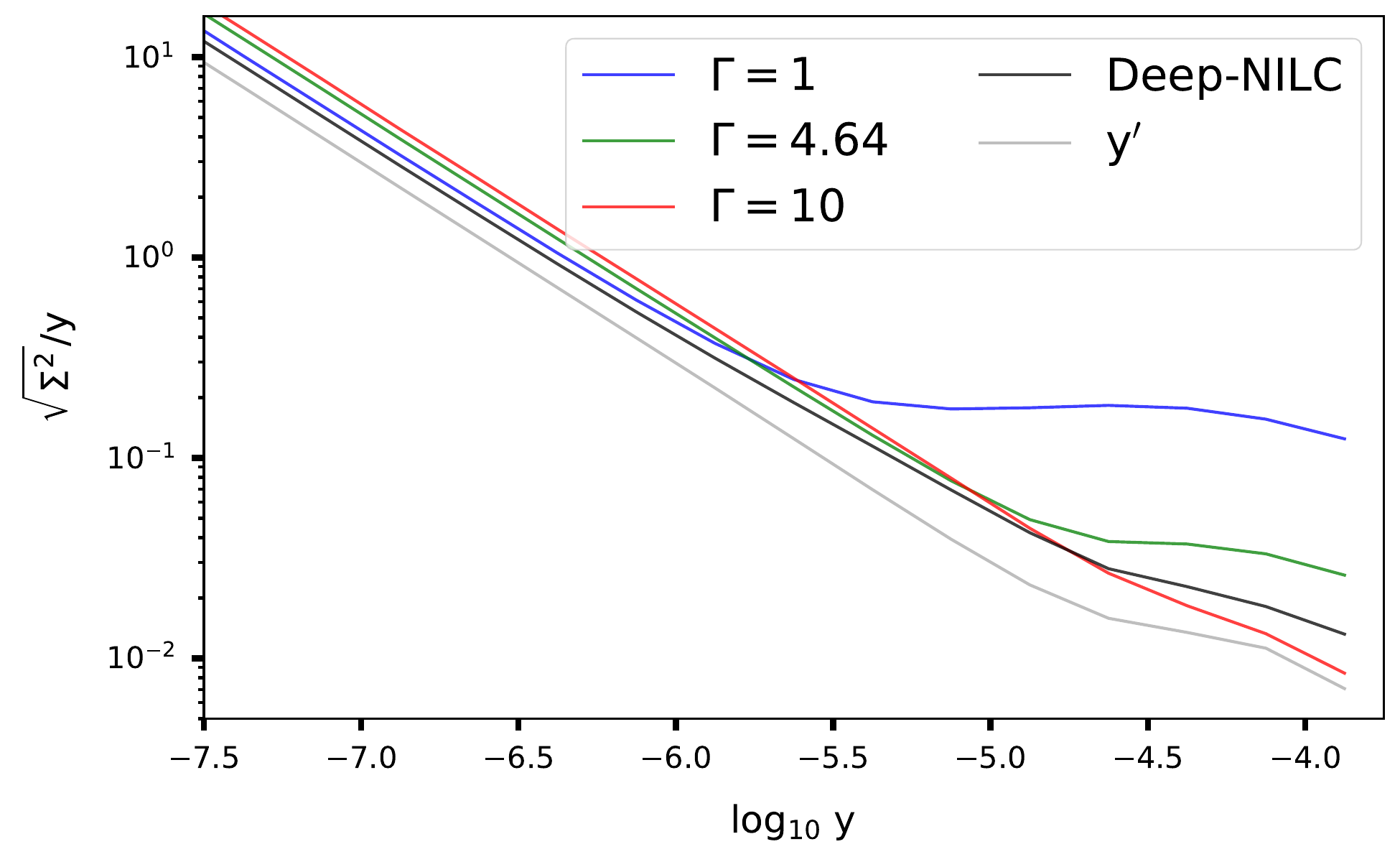}
   \includegraphics[width=0.48\textwidth]{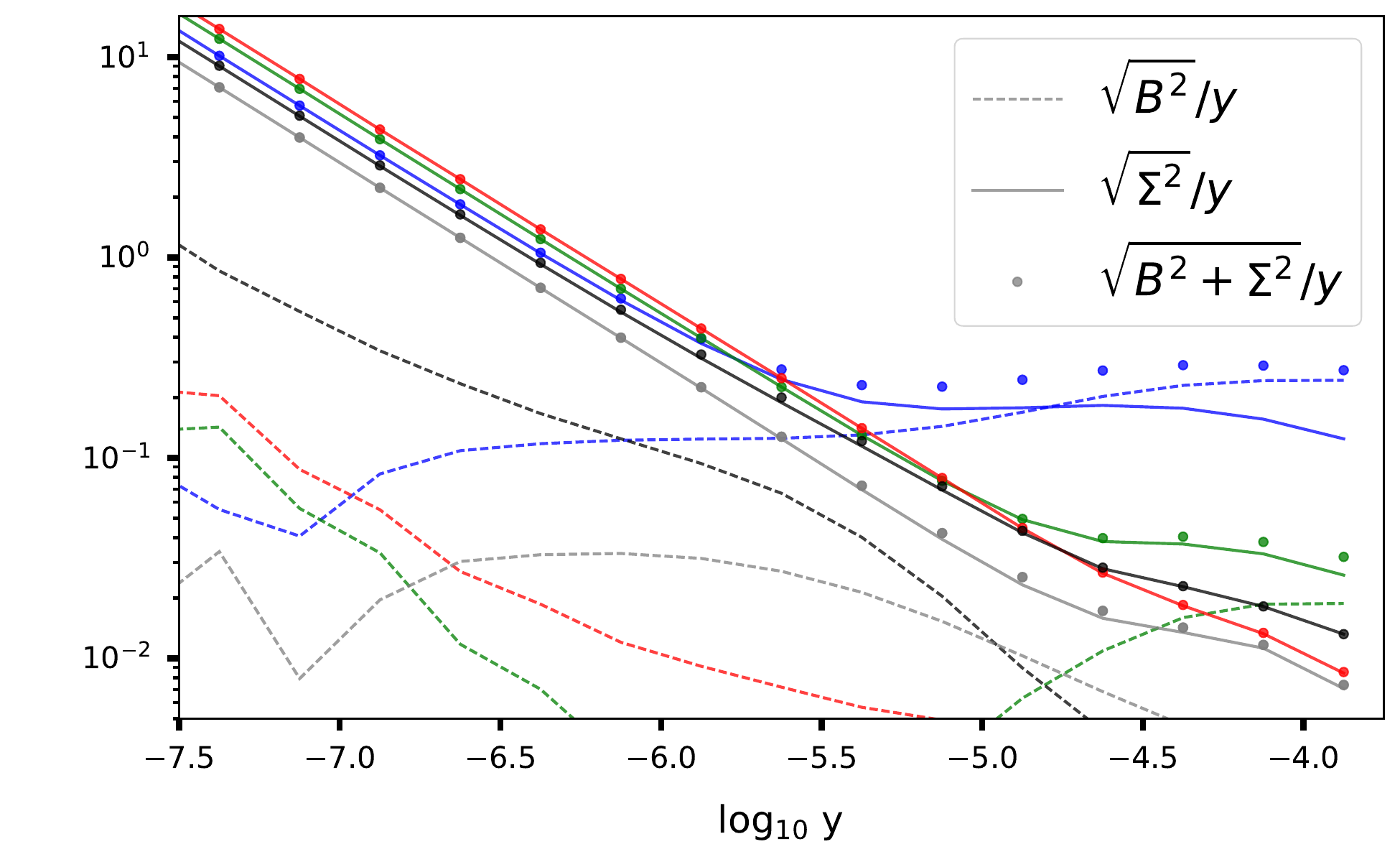}
   \includegraphics[width=0.48\textwidth]{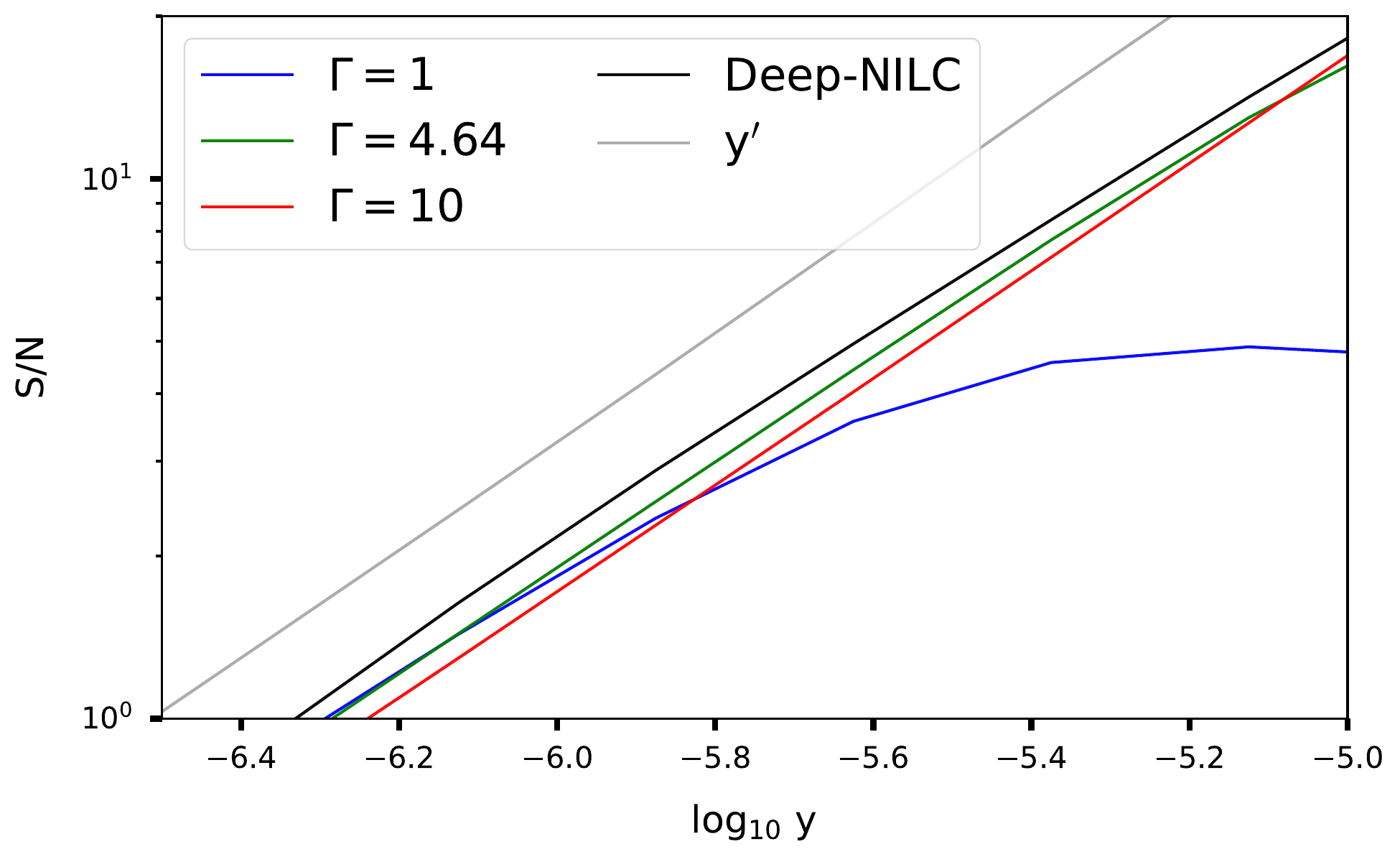}

    \caption{(top-left) The fractional bias as a function of SZ signal strength. We show the results for the three nominal values of $\Gamma$, Deep-NILC, and the best possible NILC solution. (top-right) Same as top-left panel but for the empirical scatter. (bottom-left) Displays the contributions from each source of error. The dashed lines show the absolute values of the top-left panel, solid lines show those from the top-right panel, and the circles denote the total contribution from both components. (bottom-right) The expected S/N for the pixel values. The y-axis is truncated at S/N$=1$ and where $\mathrm{log_{10}y} = -5$.}
    
\label{fig:frac_bias}
\end{figure*}

This section presents the residuals as a function of SZ signal strength. For this part, pixel values were first separated into logarithmic bins of size 0.25 dex. Then we calculated two quantities: the empirical bias $B = \langle\widehat{y}-y\rangle$ and scatter $\mathbf{\Sigma^{2} = \langle(\widehat{y}-y - B)^{2}\rangle}$. In \autoref{fig:frac_bias}, we plot the average values of the three validation maps. The color coding is the same as in \autoref{fig:residual_power}. 

In the top-left panel, we show the fractional empirical bias. The $\Gamma=1$ extraction yielded a negative fractional bias, especially for moderate to large SZ signals. Larger values of $\Gamma$ (i.e., 4.64 and 10) appeared to show almost no bias for $y>10^{-7}$, but yielded a positive bias for weaker signals. Finally, the best NILC solution showed a slight negative fractional bias that decreased as the SZ signal increased, albeit only by a few percent. The results of Deep-NILC produced an unbiased solution for strong SZ signals $y\gtrsim 10^{-5}$, however, it returned a steep negative bias moving toward weaker signals, reaching 50\% bias by $y\sim 10^{-7}$. We note the the empirical bias is not the same as the ILC bias presented in \autoref{eq:ILC_bias}. The ILC bias stems from the amount of available modes used to construct the covariance matrix in each wavelet band. The empirical bias includes the ILC bias, however, it also depends on the contaminating properties which are not stationary.

The top-right and bottom-left panels include the fractional empirical scatter. The bottom-left panel shows dashed and solid lines to denote the empirical bias and scatter respectively. The circle points represent these two quantities added in quadrature, and we call this the total fractional error. One can see the total fractional error for $\Gamma=1$ was dominated by the bias for large SZ values, but this changed as the signal dropped below $y< 10^{-5}$. For $\Gamma=10$, however, the bias remained subdominant to the scatter for all values. Looking at y$^{\prime}$, the total error was controlled by the scatter for all signal strengths. This same trend was also observed for Deep-NILC.

In the bottom-right panel we show the expected signal-to-noise which we define as
\begin{equation}
    \mathrm{S/N} = \frac{y - B}{\Sigma}
\end{equation}
The main takeaway from this plot is the S/N for Deep-NILC was larger for all fixed values of $\Gamma$ over most values of SZ signal. We discuss the implications of these results in \autoref{sec:discussion} to argue Deep-NILC offers an improved solution.

\subsection{Component Correlation}
These results show the correlation between the residual \ymaps and known sources of contamination. We used the correlation coefficient
\begin{equation}
    r_{\ell} = \frac{C_{\ell}^{a \times b}}{\sqrt{C_{\ell}^{a}C_{\ell}^{b}}}
\end{equation}
where $a$ and $b$ represent two maps of the sky, and $a \times b$ denotes a cross-correlation. We set one of these to be the residual \ymap while the other was a map of the contamination component. Most of the components did not show significant correlation with the residuals. The strongest correlations occurred with the SZ effect itself and the CIB.

\begin{figure*}
\centering
    \subfigure{\includegraphics[width=0.48\textwidth]{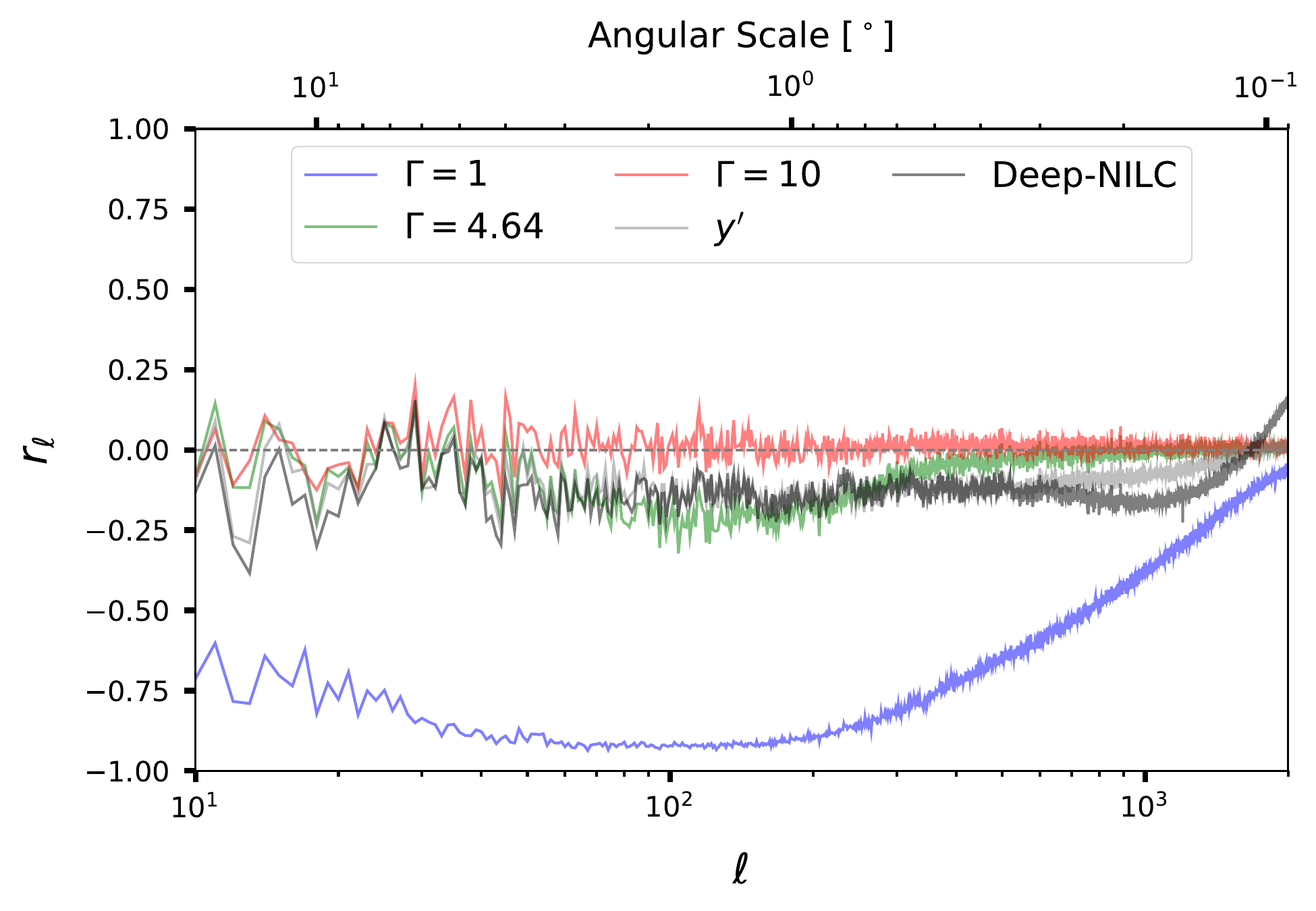}} 
    \subfigure{\includegraphics[width=0.48\textwidth]{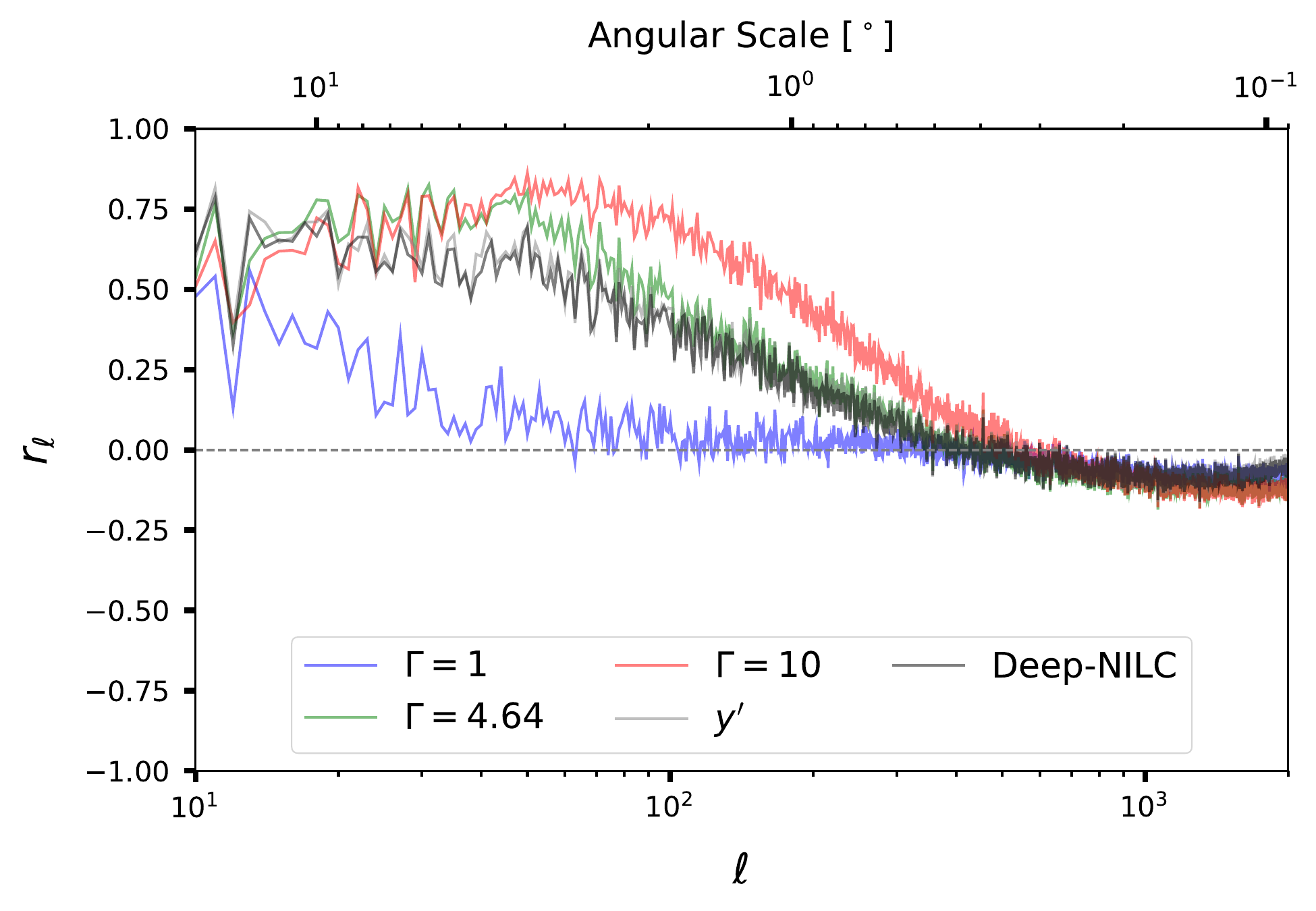}}

    \caption{Cross-correlation coefficients using the residuals from $y^{\prime}$, Deep-NILC, and the three nominal values of $\Gamma$. (left) Correlation between the residuals and the SZ signal. The best possible NILC solution is in gray and lies just below zero for all $\ell$. (right) Same as left panel but for the CIB. The main takeaway here is that Deep-NILC (black line) closely followed the best NILC solution (gray line) which was desirable.}
\label{fig:corr}
\end{figure*}

The cross-correlation coefficients between the residuals and the SZ signal are shown in the left panel of \autoref{fig:corr}. A strong negative correlation existed for $\Gamma=1$ on nearly all scales. As $\Gamma$ increased, however, the correlation coefficient decreased. Essentially, no correlation was seen by the time $\Gamma=10$. The best NILC solution also exhibited a slight negative correlation and was similar to the results of $\Gamma=4.64$ and Deep-NILC. 

The right panel of \autoref{fig:corr} shows the residual correlation with the CIB. Positive correlations were seen on moderate to large scales for most values of $\Gamma$. The most positive correlations were observed for the largest values of $\Gamma$. Deep-NILC and $y^{\prime}$ produced nearly identical correlations and were quite similar to that of $\Gamma=4.64$.

\subsection{Radial Profiles of Resolved Systems}
Here we explain how radial profiles of the resolved systems were extracted from a given \ymap. Before the extractions, the largest SZ values were masked to mitigate the contributions from nearby signals. We masked the pixels in the top 5\% of pixel values found in a $20^{\circ}$ radius in the background SZ. The group- and Coma-like signals were extracted out to 400$^{\prime}$ and 300$^{\prime}$ respectively in 10$^{\prime}$ circular bins. Mean values calculated from the outer five bins were then subtracted off as a local background correction for each field. The Virgo-like system was extracted out to 700$^{\prime}$ using the last ten annuli for the local background correction. 

Next, we calculated the stacked signal for each type of systems. For the ten Coma- and Virgo-like signals, we took the mean value and the standard deviation across each annulus as a measure of the uncertainty. We did not use the error on the mean value since these signals are unique in the sense we can only make one observation of them in the real Universe. For the galaxy group-like signal, we performed bootstrap sampling using sample sizes of ten. Galaxy group signals are too weak to be studied individually, so in practice, one would have to stack their signals to get a robust measurement. We chose a sample size of ten to mimic the sample size of nearby galaxy groups used in \citep{Pratt21}.

In the left column of \autoref{fig:stack}, we show the stacks of resolved signals. The results from Deep-NILC are plotted as black circles while those from the three nominal values of $\Gamma$ are shown as colored stars. We omit the results for $\Gamma=4.64$ in the group- and Virgo-like rows for visualization purposes; the results for $\Gamma=4.64$ generally lied between those from $\Gamma=10$ and $\Gamma=1$. The black line represents the injected profile that we tried to recover, however, it is more important to look at the cyan data. These represent the true stacked signal which is the superposition of the background SZ signals and the injected signals. In reality, one can not perfectly recover the black line as the injected signal becomes convolved with the background SZ.

For the galaxy group-like signal, the results of Deep-NILC generally were between those of $\Gamma=10$ and $\Gamma=1$. The statistical errors for Deep-NILC were roughly $60\%$ those of $\Gamma=10$, however, Deep-NILC yielded a larger bias as seen in \autoref{tab:Y}. There was very little difference between Deep-NILC and $\Gamma=10$ for the Coma-like signal, but $\Gamma=1$ performed poorly with large bias. Looking at the Virgo-like system, there was a clear negative bias for Deep-NILC while $\Gamma=10$ returned accurate and acceptable results. For better inspection, we plot the empirical bias, scatter, and total error for the stacks in the appendix.  

In the right panel of \autoref{fig:stack} we show the cumulative ``SZ flux'' given as
\begin{equation}
\label{eq:sz_flux}
    Y(r) = \sum_{i; r_{i}<r} 2\pi r_{i} y(r_{i}) \Delta r
\end{equation}
each radial bin is indexed by $i$, and $\Delta r = 10^{\prime}$ which is the angular width of each bin. This observed quantity is important as it is directly relatable to the total gas mass enclosed within some radius. We estimated the SZ flux values within 1, 2, and 3 R$_{500}$ for each system indicated by the dotted brown lines. Then we calculated the percent differences from the true Y for the three $\Gamma$ values and Deep-NILC which are presented in \autoref{tab:Y}. We discuss these results further in \autoref{sec:discussion}.

\begin{table*}
\centering

\caption{Percent Difference in Y}
\centering
\begin{tabular}{|c||ccc|ccc|ccc|}

\hline
Galaxy Group & {} & R$_{500}$ &{} & {}& 2 R$_{500}$&{}& {}& 3 R$_{500}$ &{}\\
{} & $\langle B \rangle$ & $\langle \Sigma \rangle$ &$\langle \sqrt{B^{2}+\Sigma^{2}} \rangle$ & $\langle B \rangle$ & $\langle \Sigma \rangle$ &$\langle \sqrt{B^{2}+\Sigma^{2}} \rangle$& $\langle B \rangle$ & $\langle \Sigma \rangle$ &$\langle \sqrt{B^{2}+\Sigma^{2}} \rangle$\\
\hline
$\Gamma = 10$ & -18.6\% & 51.1\% & 54.4\% & -14.8\% & 62.2\% & 64.0\% & 35.7\% & 76.3\% & 84.3\%\\
$\Gamma = 4.64$ & -15.9\% & 44.0\% & 46.7\% & 6.66\% & 53.2\% & 53.6\% & 16.9\% & 69.2\% & 71.2\%\\
$\Gamma = 1$ & -55.2\% & 21.0\% & 59.0\% & -60.2\% & 22.7\% & 64.4\% & -72.2\% & 36.3\% & 80.9\%\\
Deep-NILC & -42.3\% & 28.7\% & 51.1\% & -41.0\% & 24.4\% & 47.7\% & -39.3\% & 33.1\% & 51.4\%\\
\hline
\hline

Coma & {} & R$_{500}$ &{} & {}& 2 R$_{500}$&{}& {}& 3 R$_{500}$ &{}\\
{} & $\langle B \rangle$ & $\langle \Sigma \rangle$ &$\langle \sqrt{B^{2}+\Sigma^{2}} \rangle$ & $\langle B \rangle$ & $\langle \Sigma \rangle$ &$\langle \sqrt{B^{2}+\Sigma^{2}} \rangle$& $\langle B \rangle$ & $\langle \Sigma \rangle$ &$\langle \sqrt{B^{2}+\Sigma^{2}} \rangle$\\
\hline
$\Gamma = 10$ & 0.6\% & 1.1\% & 1.3\% & 1.6\% & 2.5\% & 2.9\% & 2.5\% & 4.5\% & 5.1\%\\
$\Gamma = 4.64$ & -8.4\% & 4.1\% & 9.3\% & -5.2\% & 3.2\% & 6.1\% & -1.9\% & 4.9\% & 5.3\%\\
$\Gamma = 1$ & -79.2\% & 7.0\% & 79.5\% & -75.4\% & 8.2\% & 75.8\% & -65.4\% & 11.5\% & 66.4\%\\
Deep-NILC & 0.4\% & 1.9\% & 1.9\% & 0.8\% & 2.3\% & 2.4\% & 0.8\% & 4.1\% & 4.2\%\\
\hline
\hline

Virgo & {} & R$_{500}$ &{} & {}& 2 R$_{500}$&{}& {}& 3 R$_{500}$ &{}\\
{} & $\langle B \rangle$ & $\langle \Sigma \rangle$ &$\langle \sqrt{B^{2}+\Sigma^{2}} \rangle$ & $\langle B \rangle$ & $\langle \Sigma \rangle$ &$\langle \sqrt{B^{2}+\Sigma^{2}} \rangle$& $\langle B \rangle$ & $\langle \Sigma \rangle$ &$\langle \sqrt{B^{2}+\Sigma^{2}} \rangle$\\
\hline
$\Gamma = 10$ & 2.4\% & 2.7\% & 3.6\% & 2.4\% & 4.2\% & 4.8\% & 2.3\% & 8.5\% & 8.8\%\\
$\Gamma = 4.64$ & 1.1\% & 2.1\% & 2.4\% & 3.0\% & 3.8\% & 4.8\% & 3.5\% & 7.1\% & 7.9\%\\
$\Gamma = 1$ & -51.4\% & 10.3\% & 52.4\% & -42.9\% & 10.5\% & 44.1\% & -41.0\% & 14.2\% & 43.4\%\\
Deep-NILC & -2.6\% & 3.4\% & 4.3\% & -4.5\% & 5.3\% & 7.0\% & -7.4\% & 7.4\% & 10.5\%\\
\hline

\label{tab:Y}
\end{tabular}
\end{table*}

\section{Discussion} \label{sec:discussion}

\subsection{Deep-NILC vs. Fixed Spatial Parameter}
Our results demonstrated that Deep-NILC outperformed NILC algorithms that assumed a fixed value of $\Gamma$. Small values of $\Gamma$, specifically $\Gamma=1$, produced highly biased results. This was most evident in the top-left panel \autoref{fig:frac_bias} where the average bias was $\sim 10\%-20\%$. It was also manifested in the bottom-left panel of \autoref{fig:frac_bias} where the bias term provided the most contribution to the total fractional scatter for large SZ signals. The same idea was also reflected in the left panel of \autoref{fig:corr} which showed a very strong correlation between the residuals and the SZ signal. This is why $\Gamma=1$ exhibited the largest residual power in \autoref{fig:residual_power}. 

Furthermore, $\Gamma=1$ yielded the worst extraction for Coma- and Virgo-like systems, however, its performance was better for the galaxy group-like signal. For the pixel-by-pixel statistics, $\Gamma=1$ yielded the best results for weaker SZ signals and on the largest scales. Part of the reason could be due to the lack of correlation seen between the residuals and the CIB. In other words, smaller values $\Gamma$ did better at extracting weak and/or large-scale signals because they reduced the effects from contamination. Nevertheless, the large observed bias was prominent and would not be suitable for most SZ studies.

The story was essentially reversed when looking at larger values of $\Gamma$. We found that $\Gamma=10$ provided the most unbiased solutions but with the strongest residuals. In \autoref{fig:frac_bias} we demonstrated the bias term was insignificant compared to the empirical scatter. Moreover, there was a modest positive bias for weak SZ signals at large radii which likely coincided with the positive correlation seen with the CIB in \autoref{fig:corr}. 

\begin{figure*}
\centering
    \subfigure{\includegraphics[width=0.48\textwidth]{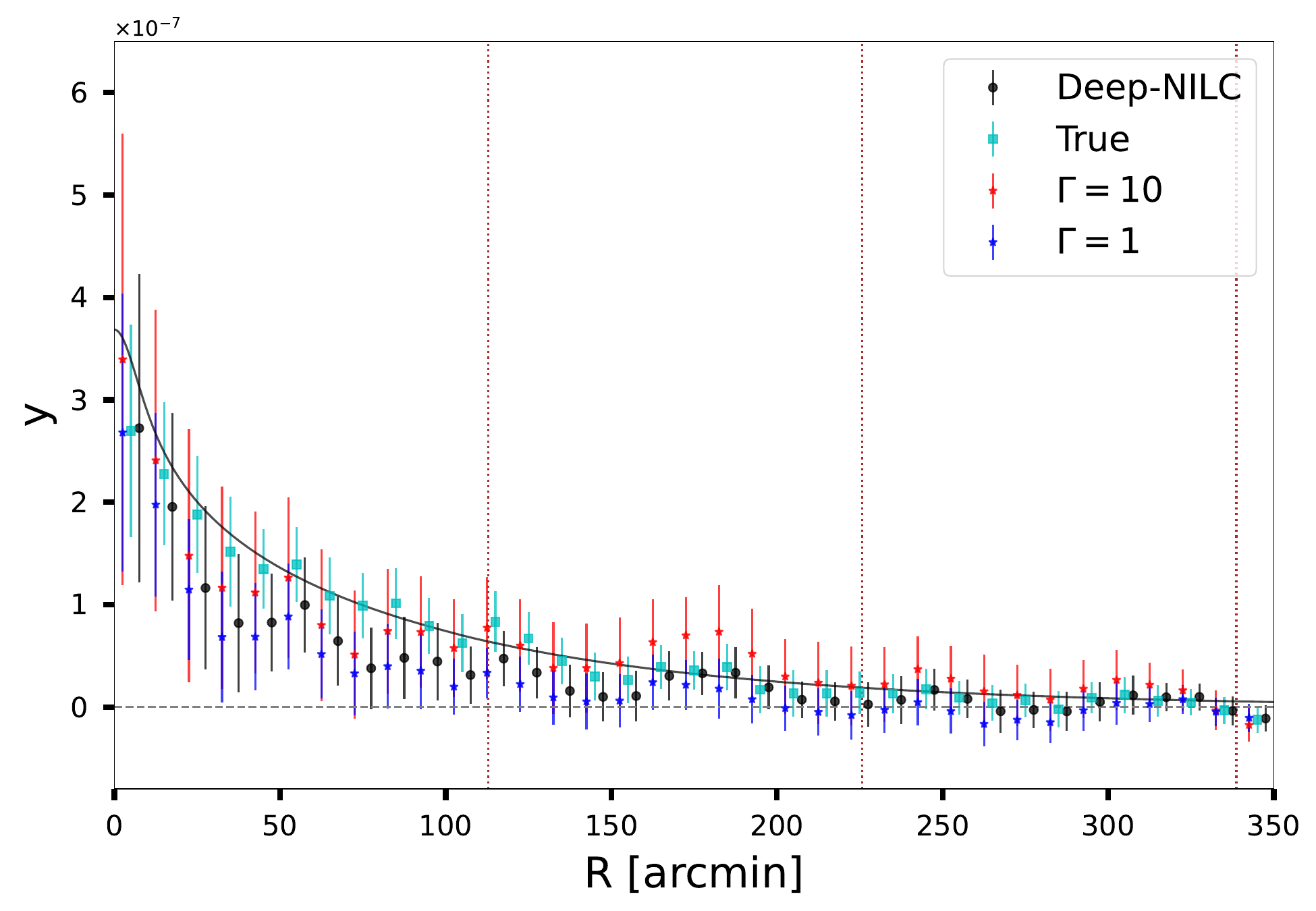}}
    \subfigure{\includegraphics[width=0.48\textwidth]{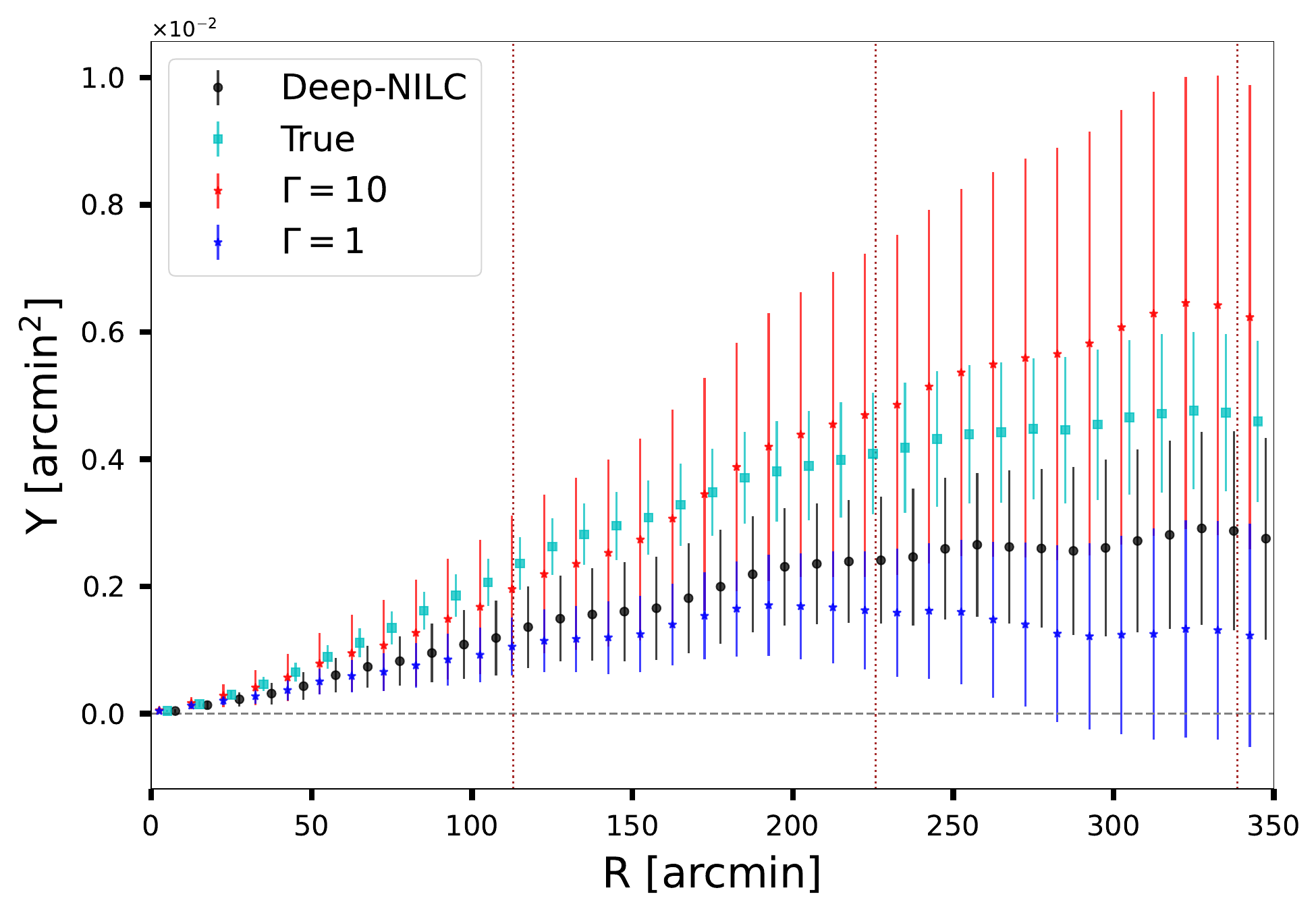}}
    \subfigure{\includegraphics[width=0.48\textwidth]{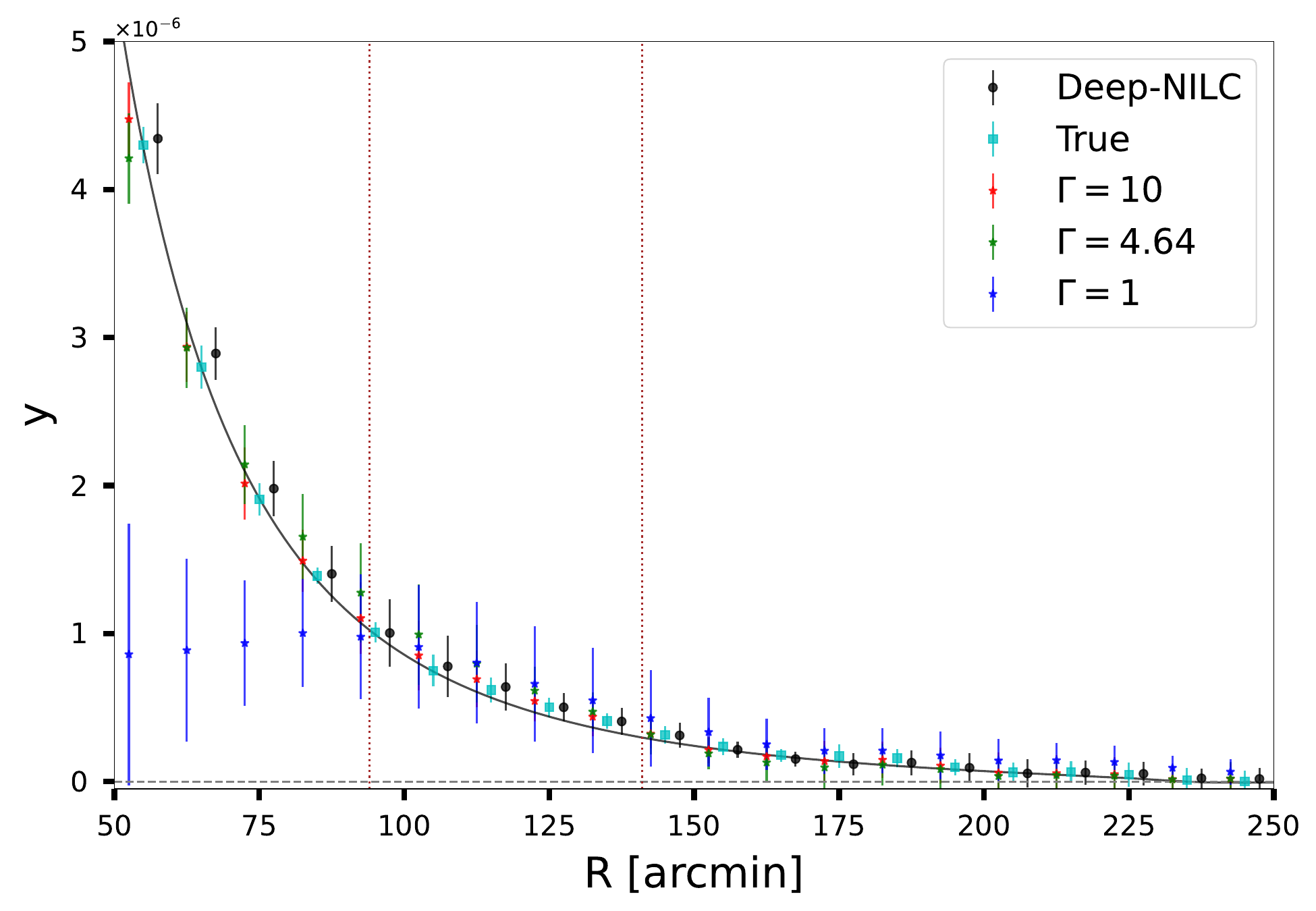}}
    \subfigure{\includegraphics[width=0.48\textwidth]{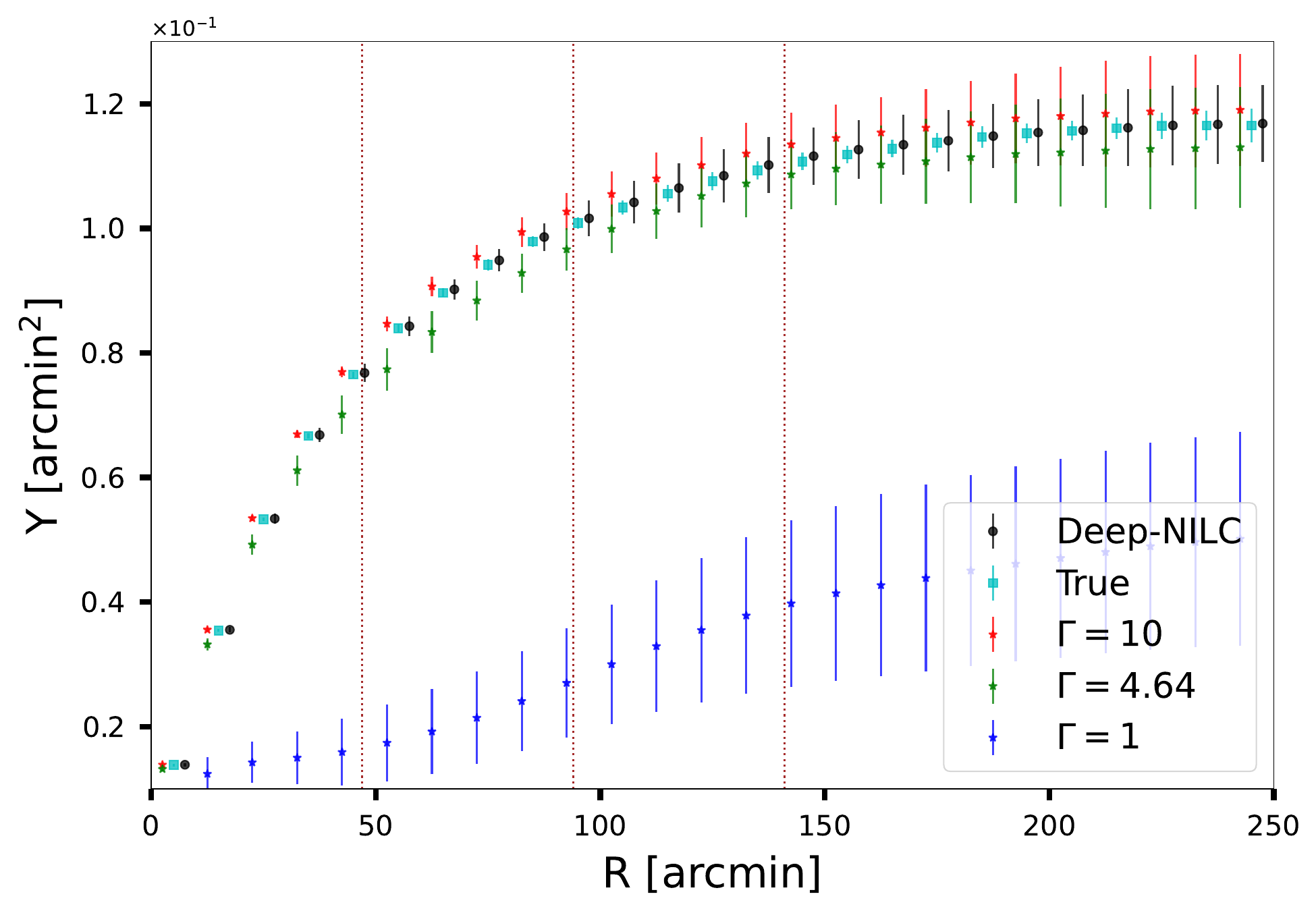}}
    \subfigure{\includegraphics[width=0.48\textwidth]{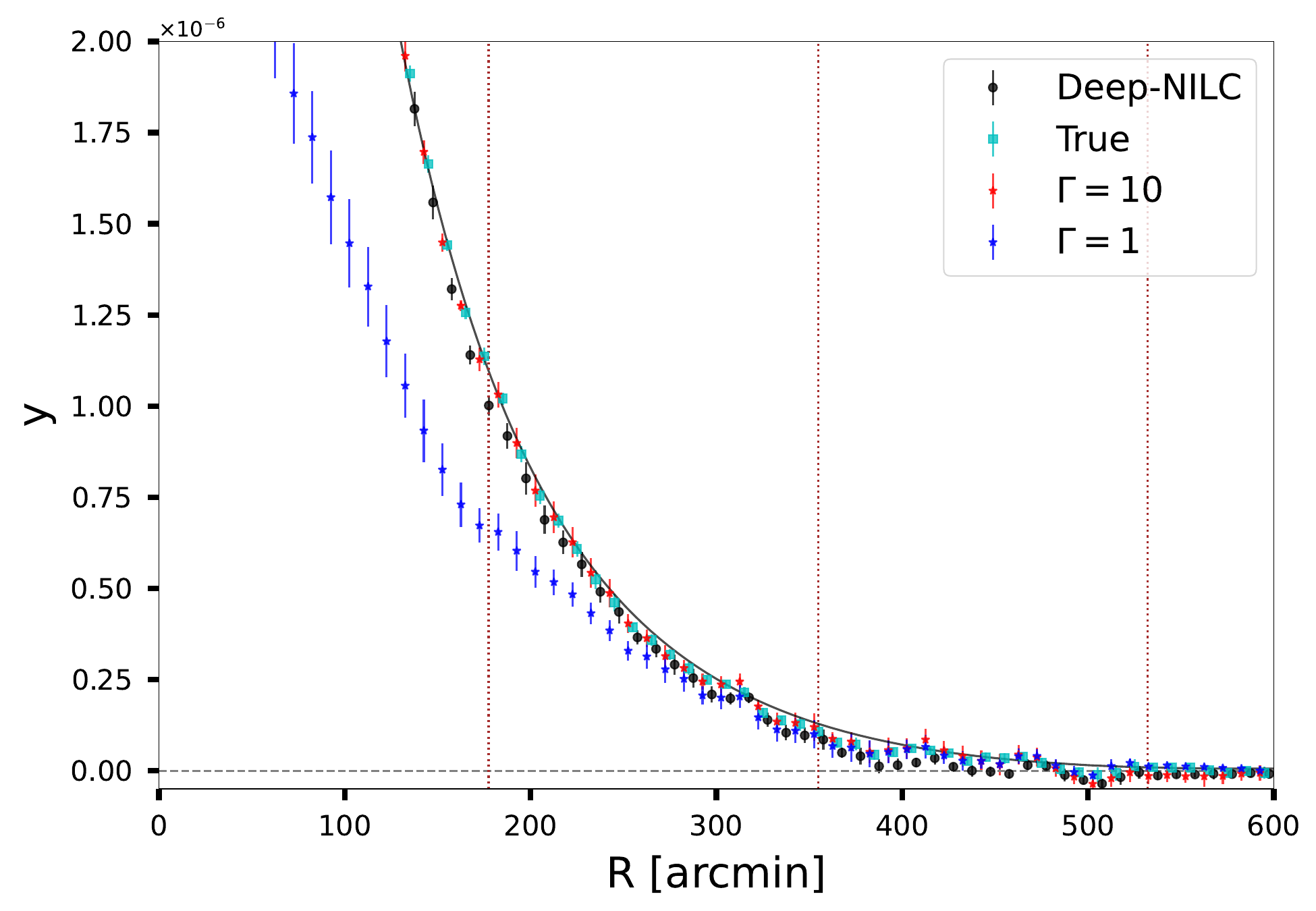}}
    \subfigure{\includegraphics[width=0.48\textwidth]{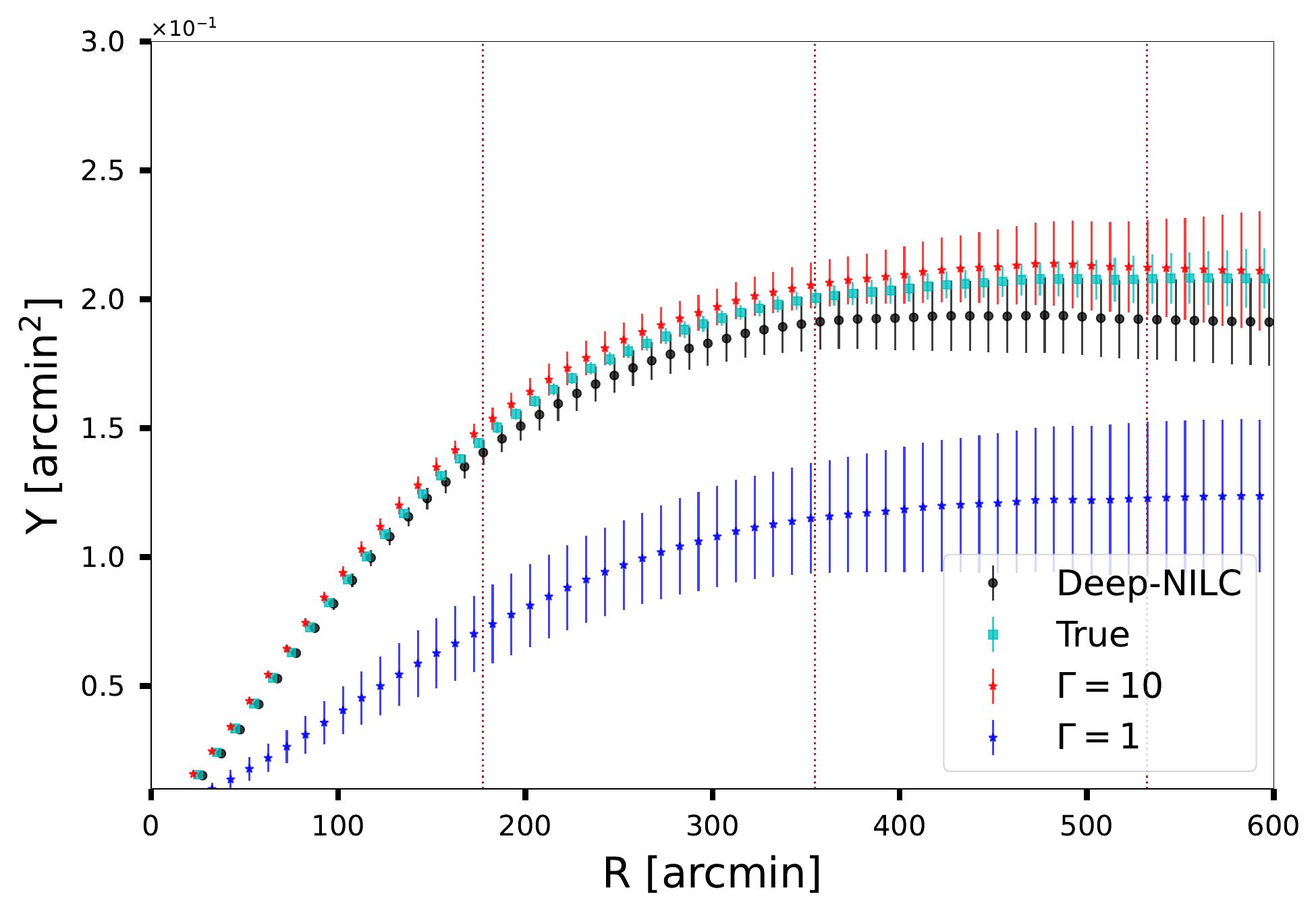}}

    \caption{The left column shows the stack of ten radial profiles for three resolved signals. The right column shows the cumulative SZ flux. The group-, Coma-, and Virgo-like signals are shown in the top, middle and bottom panels respectively. The results from different values of $\Gamma$ are shown as stars, those from Deep-NILC as black circles, and the true signal as cyan squares. The stars and circles are slightly offset in the horizontal direction to help with visualization. Extractions for $\Gamma=4.64$ are not shown in the group- and Coma-like panels to avoid over-crowding. The black line shows the profile that was injected into the simulated \ymap, and the dotted brown lines indicate the positions of 1, 2, and 3 R$_{500}$. The error bars show the standard deviation of the mean for each annulus. The cyan points do not exactly match the input profile since the injected signal has been convolved with other nearby SZ signals. The error bars on the cyan squares are the statistical fluctuations of the true signal within the stack after some masking. The main point of this plot is to show how well resolved signals are recovered with Deep-NILC and different values of $\Gamma$.}
\label{fig:stack}
\end{figure*}

Deep-NILC outperformed the extractions for both large and small values of $\Gamma$ by providing a balance between the bias and variance. This balance was seen in the bottom-left panel of \autoref{fig:frac_bias} where Deep-NILC approached the solution for $\Gamma=10$ for large SZ values and that of $\Gamma=1$ for weak signals. These results mean that Deep-NILC was able to adjust the localization based on the strength of the SZ signal. We also observed Deep-NILC to have the smallest residual power in \autoref{fig:residual_power} and more accurate recoveries of the group- and Coma-like profiles in \autoref{fig:stack} and \autoref{tab:Y}.

\subsection{Caveats}
Deep-NILC comes with limitations. Most importantly, the algorithm relies heavily on the quality of simulated data. The data used in this work were generated using a set of ten different \ymaps but only one realization of contaminating components. One might argue that this could lead to over-fitting to this specific set of signals. In the future, emphasis should be placed on including varying the foreground emission models. However, we note that the NILC estimate was also affected by other nearby SZ signals as they were contained in the covariance matrix. In this sense, the SZ signal can be thought of a contaminant itself using NILC. One way around this might be to mask strong SZ sources before constructing $\mathbf{R}$. Nevertheless, the different \ymaps we used gave a variety SZ distributions, and this helped provide generalization to our model.

Also, building a model with supervised learning is sensitive to the diversity of training examples. We had to include additional resolved SZ signals to augment the \ymaps from \citet{Han21}. This was imperative for our model to learn about the resolved ``anomalies'' seen in our real sky, such as the Coma and Virgo clusters. It is likely, however, that there were not enough examples in the training data of a Virgo-like cluster. The Virgo cluster is very unique as it is the only prominent SZ signal that extends multiple degrees on the sky. In our methodology, we injected resolved systems based on the MCXC catalog, so only a few instances of Virgo-like systems were included in the training set. In \autoref{fig:stack}, a non-trivial bias was observed for the Virgo-like stack, however, this was not observed for the Coma-like and galaxy group-like signals which were much smaller in angular size. Furthermore, the majority of SZ sources produced zero signal on large-scales. Wrapping this together with the sample imbalance is what probably caused Deep-NILC to under-predict the signal of Virgo.

Another potential drawback of our simulated data set was the omission of spatially correlated components. For example, a background AGN may produce strong emissions both in radio and IR. If strong enough, they may significantly affect the NILC extraction. Also, radio-loud AGNs often sit at the center of galaxy clusters (e.g., Virgo cluster). Deep-NILC was not trained to see these spatially correlated components, thus, we caution that Deep-NILC could yield unreliable results in some instances of real data. This may be worth exploring in future work as simulations improve.

The simulated data used in this work also did not consider instrumental noise or beam effects. Currently, adding realistic instrumental noise is not incorporated into PySM, however, one could simply add white Gaussian noise to the frequency data. Adding instrumental noise would affect the calculation of the ILC weights and raise the reconstructed uncertainty floor. In turn, the gray line in \autoref{fig:residual_power} would shift upward along with the rest of the lines. Deep-NILC, however, does not explicitly handle the noise after the ILC calculation as it takes the different NILC solutions as preprocessed input. In addition, beam effects were ignored in the development of Deep-NILC. For real data, one must deconvolve the individual frequency maps and reconvolve them to a common frequency \citep[e.g.,]{McCarthy23,Chandran23}. We do not include this component to focus on the method as a proof-of-concept.

Finally, we reiterate that Deep-NILC was still bounded by the capabilities of the NILC algorithm. This is because we regressed using the best NILC solution, so Deep-NILC could not perform better than the $\Gamma$ values used here. If no combination of the $\Gamma$ values could yield reliable results then Deep-NILC would not do any better. Furthermore, we did not investigate the full parameter space of NILC. Even though we considered a grid of spatial parameters, we chose to fix the set of harmonic window functions. Using a more diverse set of harmonic filters could help improve the performance of Deep-NILC, but we leave this to future work.

\section{Summary} \label{sec:summary}
This study investigated the systematic effects of the NILC method when extracting the SZ signal. Our methods included injecting synthetically generated \ymaps into a simulated foreground of radio/IR emissions. These emissions were then computed as band-averaged signals that would mimic the observations seen by the \Planck and \WMAP satellites. 

Next, we performed the NILC component separation technique in attempt to recover the input SZ signal. The main focus was on the spatial parameter, $\Gamma$, which controlled the amount of contamination and bias. Mathematically, we demonstrated an optimal value for $\Gamma$ exists that minimizes the contribution from both of these effects. It was then argued that it is nearly impossible to analytically calculate the best value of $\Gamma$ {\it{a priori}} as it depended on many complicated variables, such as the varying level of nearby contamination. 

We went on to show how altering the value of $\Gamma$ significantly changed the performance of the SZ extraction. Specifically, we found that small values of $\Gamma$ tend to yield better results when attempting to extract weak SZ features and/or when there exists significant contributions from contamination. One the other hand, bigger values of $\Gamma$ yielded unbiased solutions and performed better when extracting large SZ signals. 

Rather than using a fixed value of $\Gamma$, we developed an algorithm to improve the NILC solution known as Deep-NILC. Deep-NILC required various NILC extractions for different values of $\Gamma$. These were then fed into a MLP to predict the SZ signal. The main takeaway from our experiments are the following:
\begin{itemize}
    \item Deep-NILC yielded smaller residual power on all angular scales compared to all values of fixed $\Gamma$ as shown in \autoref{fig:residual_power}.
    \item The correlation between the Deep-NILC residuals and known contaminants, particularly the CIB and the SZ signal itself, were closest to that obtained by the best possible NILC solution (see \autoref{fig:corr}). 
    \item The sum of the empirical bias and scatter terms seen in \autoref{fig:frac_bias} suggested Deep-NILC rendered the smallest deviations for nearly all strengths of SZ signals. The one exception being at $y\gtrsim 10^{-4}$ where $\Gamma=10$ performs slightly better, but this was a marginal difference.
    \item When stacking the resolved profiles for galaxy group-like signal, Deep-NILC yielded the best results when considering the total contributions from the bias and scatter.
    \item Deep-NILC under-predicted the signal coming from a Virgo-like source which was likely due to an imbalance in the training data. However, the inferred SZ flux for Deep-NILC was only biased at the 10\% level.
\end{itemize}

There are many other approaches that can, and should, be investigated using machine learning. As simulation data quality continues to improve and as future missions bring observational insights, data-driven models may soon supersede analytical techniques. Deep-NILC provides a bridge between analytical and data-driven solutions which could provide potential improvements to a variety of problems. In the future, one might be able to build a model that could learn the SZ signal without doing any pre-processed component separation.

We would like to thank the anonymous referee for providing excellent feedback to help improve this work. We would also like to thank Camille Avestruz, Jennier Li, and Ashley Villar for useful conversations regarding DL. In addition, we want to thank Charles Antonelli and John Theils for computational support. This project also made use of the Great Lakes computing cluster at the University of Michigan\textendash Ann Arbor. We are grateful for support from the NASA grants AWD012791 and 21-ADAP21-0069.

\clearpage
\appendix
\section{Wavelet Bands}
In the figure below, we show the wavelet bands in multipole space. The black lines are show the individual bands while the dashed red line is the full beam profile. The full profile is a Gaussian beam with FWHM = 10$^{\prime}$ truncated at FWHM = 20$^{\circ}$.
\begin{figure*}[!h]
\centering
\includegraphics[width=0.48\textwidth]{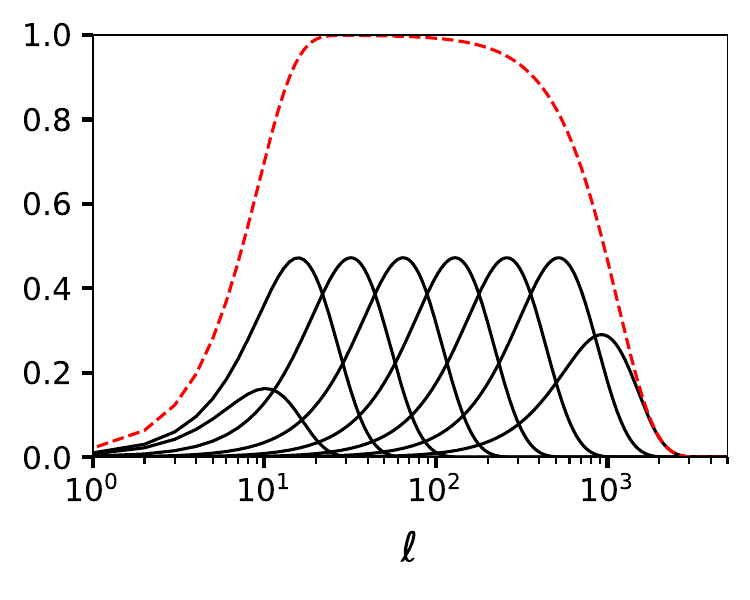}
\label{fig:wavelet}
\end{figure*}

\section{Bias and Scatter of Radial Profiles}
\begin{figure*}[!h]
    \centering
    \includegraphics[width=0.48\textwidth]{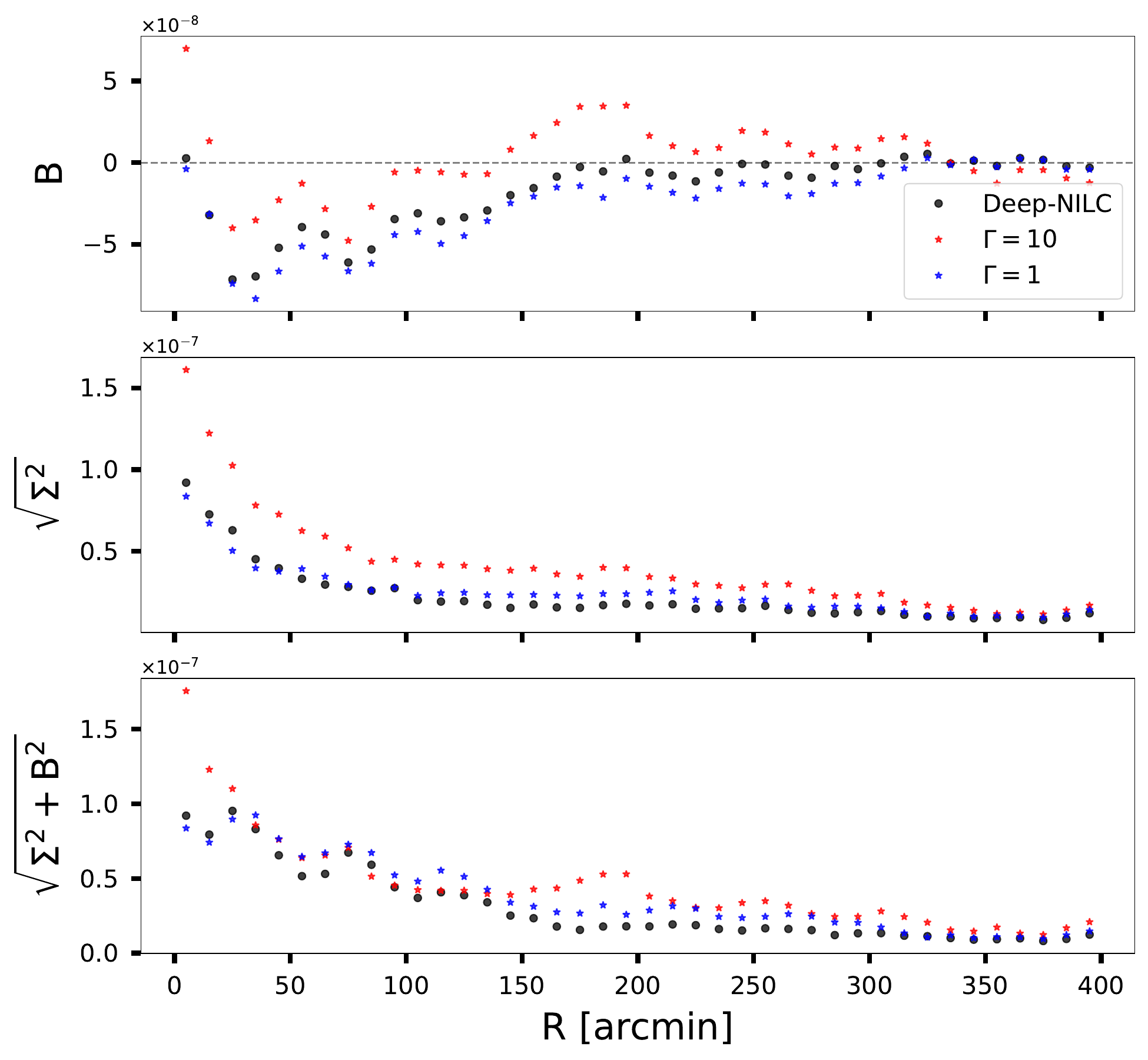}
    \includegraphics[width=0.48\textwidth]{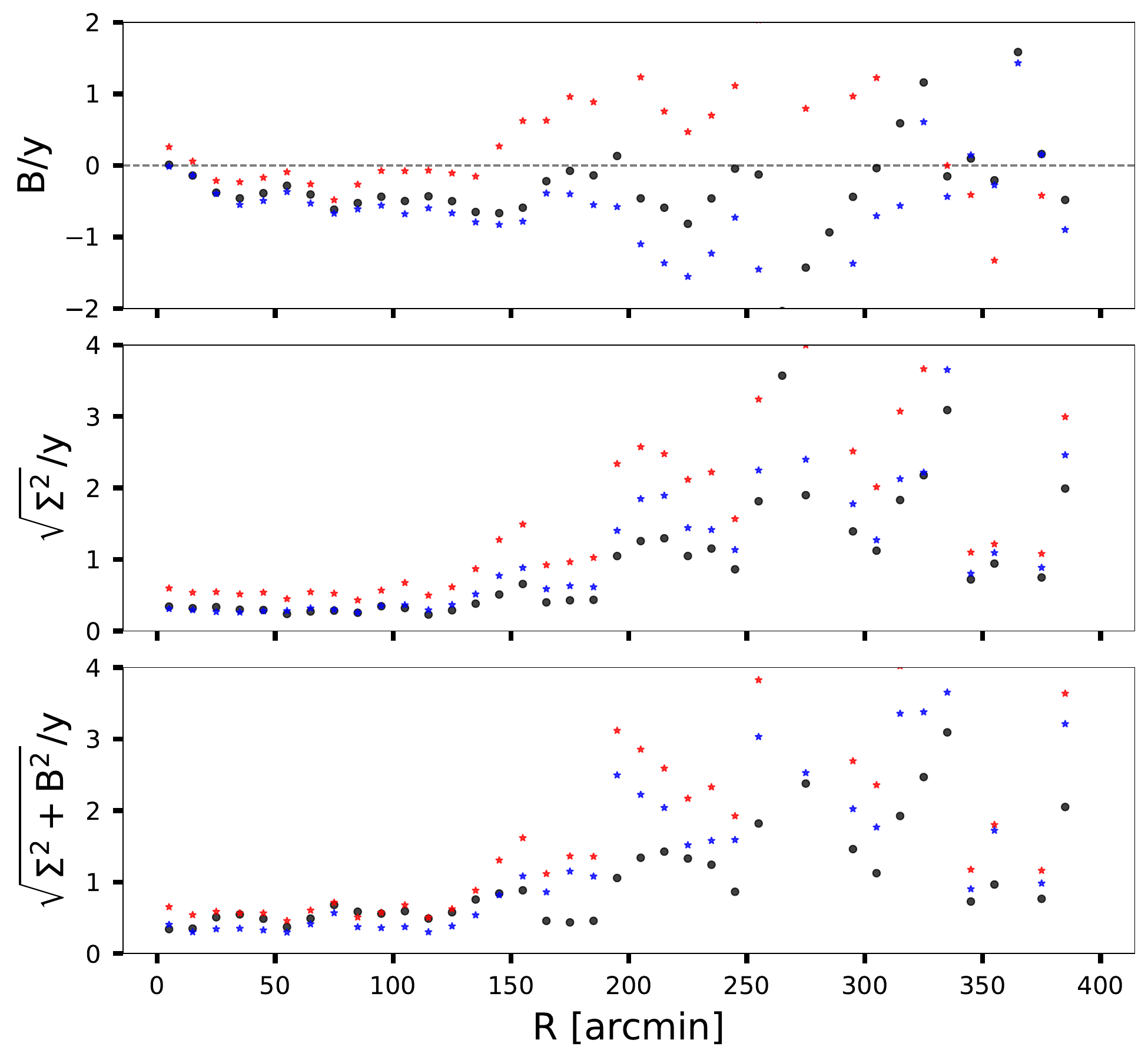}
    \caption{The average bias, scatter, and total error for galaxy group-like signals are shown in the top, middle, and bottom panels respectively. These quantities are shown in the left column while they are shown as a fraction of the SZ signal in the right column.}
    \label{fig:group_stats}
\end{figure*}

\begin{figure*}
    \centering
    \includegraphics[width=0.48\textwidth]{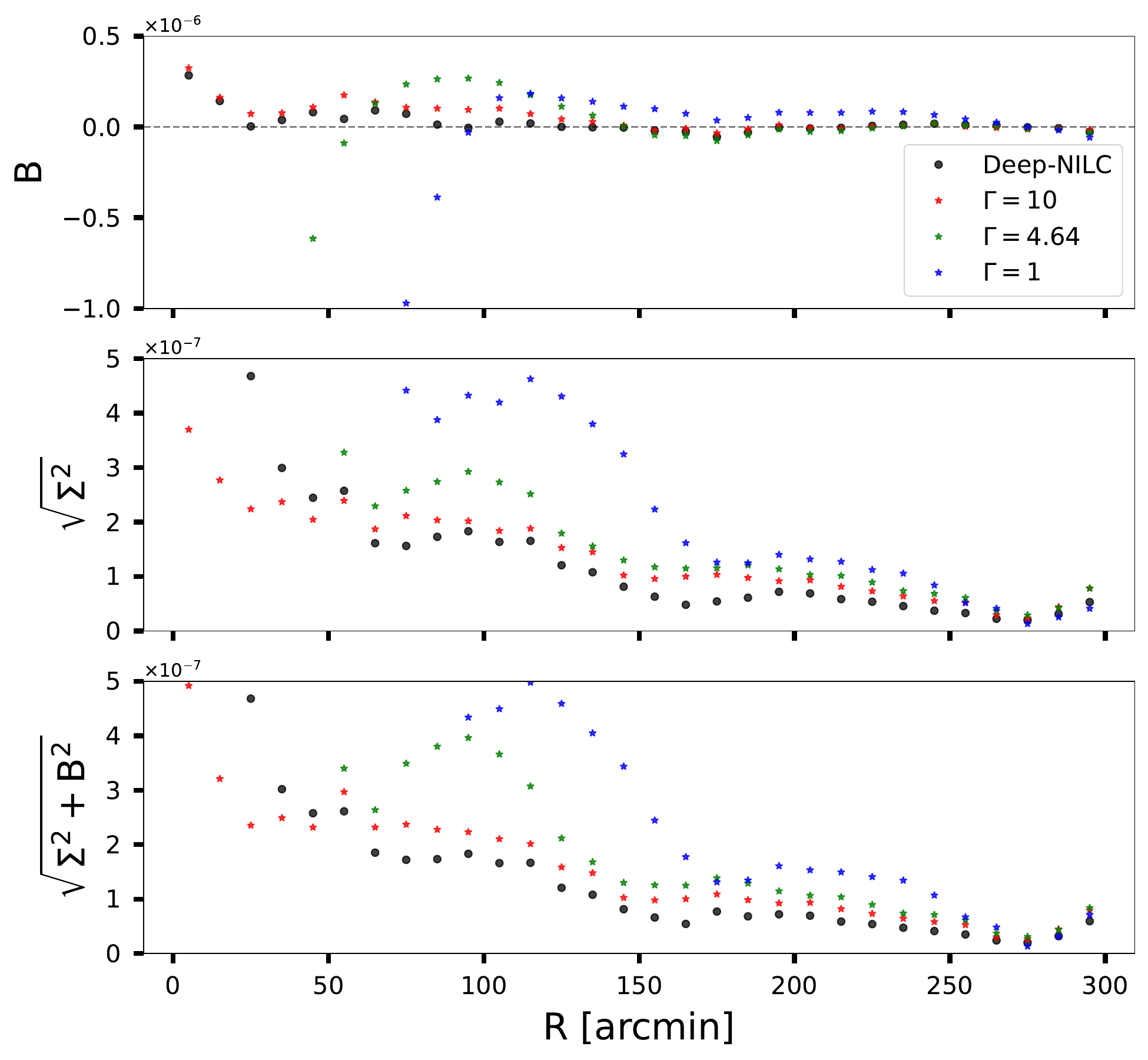}
    \includegraphics[width=0.48\textwidth]{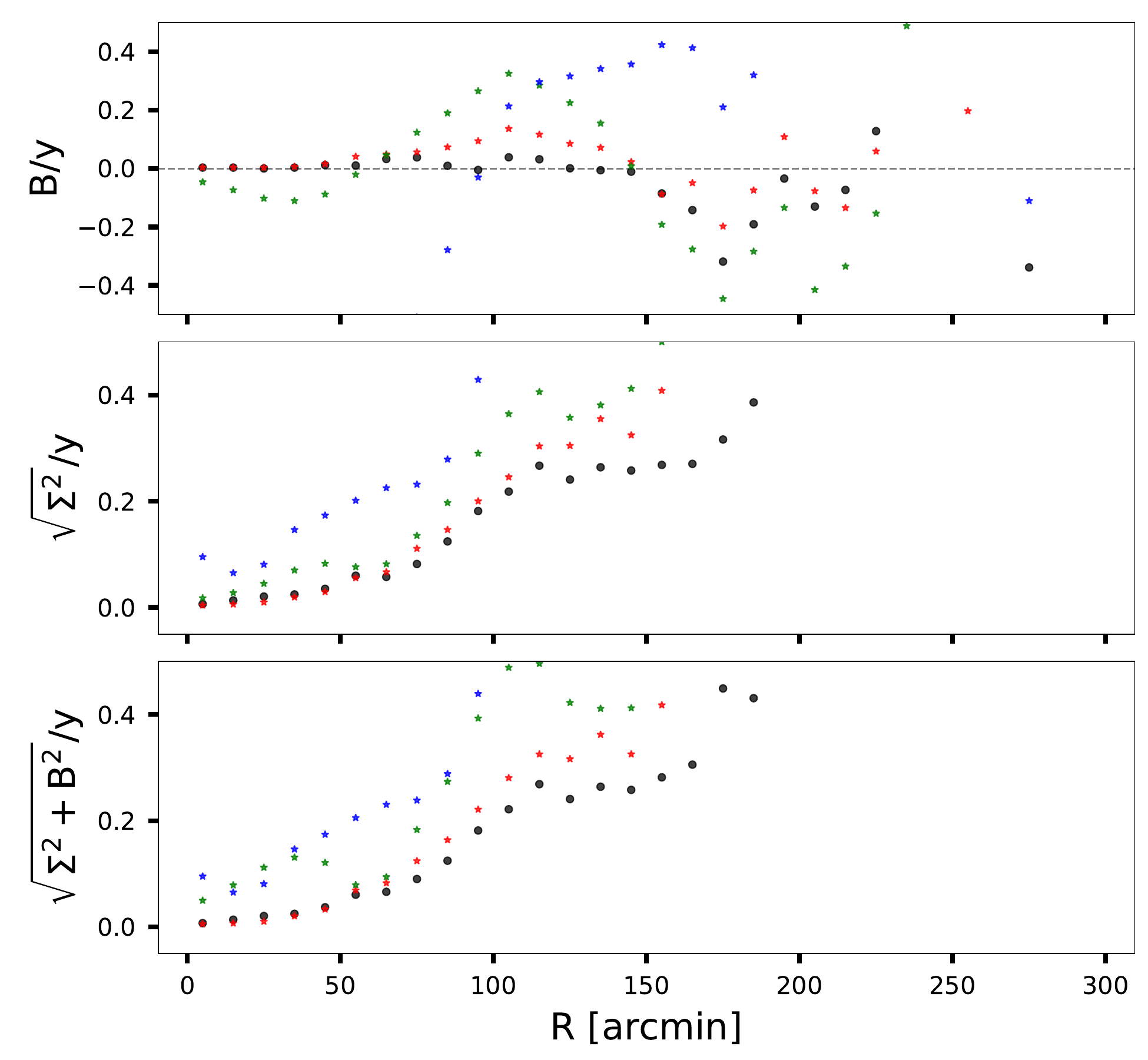}
    \caption{Same as \autoref{fig:group_stats} but for a Coma-like system.}
    \label{fig:coma_stats}
\end{figure*}

\begin{figure*}
    \centering
    \includegraphics[width=0.48\textwidth]{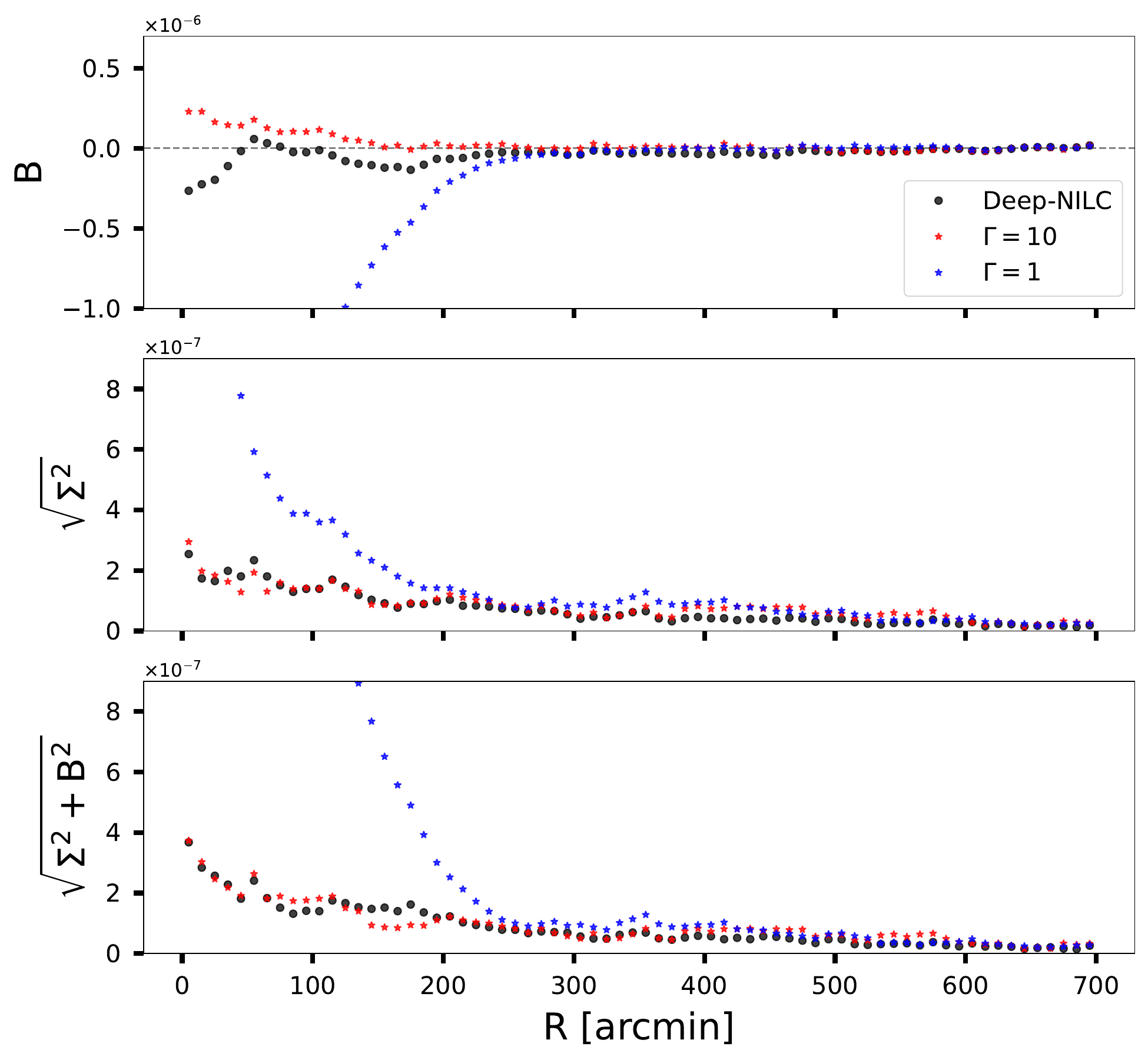}
    \includegraphics[width=0.48\textwidth]{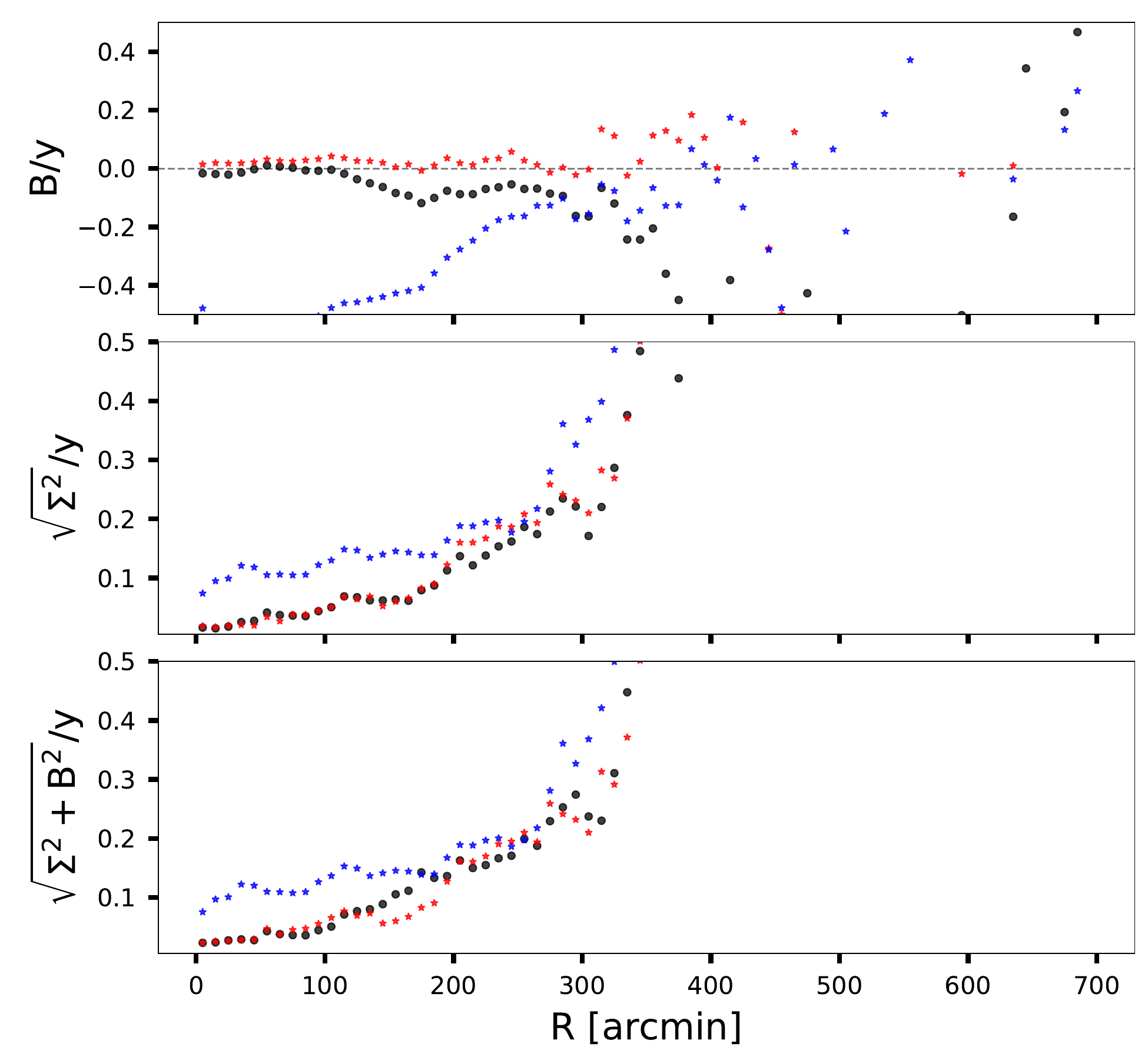}
    \caption{Same as \autoref{fig:group_stats} but for a Virgo-like system.}
    \label{fig:virgo_stats}
\end{figure*}

Here we provide the empirical bias, scatter, and total error for the resolved systems considered in this work: galaxy group-, Coma-, and Virgo-like systems shown in \autoref{fig:group_stats}, \autoref{fig:coma_stats}, and \autoref{fig:virgo_stats}.

\clearpage

\bibliography{bib.bib}

\begin{thebibliography}{}
\expandafter\ifx\csname natexlab\endcsname\relax\def\natexlab#1{#1}\fi
\providecommand{\url}[1]{\href{#1}{#1}}
\providecommand{\dodoi}[1]{doi:~\href{http://doi.org/#1}{\nolinkurl{#1}}}
\providecommand{\doeprint}[1]{\href{http://ascl.net/#1}{\nolinkurl{http://ascl.net/#1}}}
\providecommand{\doarXiv}[1]{\href{https://arxiv.org/abs/#1}{\nolinkurl{https://arxiv.org/abs/#1}}}

\bibitem[{{Bobin} {et~al.}(2008){Bobin}, {Moudden}, {Starck}, {Fadili}, \& {Aghanim}}]{Bobin08}
{Bobin}, J., {Moudden}, Y., {Starck}, J.~L., {Fadili}, J., \& {Aghanim}, N. 2008, Statistical Methodology, 5, 307, \dodoi{10.1016/j.stamet.2007.10.003}

\bibitem[{{Bobin} {et~al.}(2015){Bobin}, {Sureau}, \& {Starck}}]{Bobin15}
{Bobin}, J., {Sureau}, F., \& {Starck}, J.-L. 2015, arXiv e-prints, arXiv:1508.07131, \dodoi{10.48550/arXiv.1508.07131}

\bibitem[{{Bregman} {et~al.}(2018){Bregman}, {Anderson}, {Miller}, {Hodges-Kluck}, {Dai}, {Li}, {Li}, \& {Qu}}]{Bregman18}
{Bregman}, J.~N., {Anderson}, M.~E., {Miller}, M.~J., {et~al.} 2018, \apj, 862, 3, \dodoi{10.3847/1538-4357/aacafe}

\bibitem[{{Bregman} {et~al.}(2022){Bregman}, {Hodges-Kluck}, {Qu}, {Pratt}, {Li}, \& {Yun}}]{Bregman22}
{Bregman}, J.~N., {Hodges-Kluck}, E., {Qu}, Z., {et~al.} 2022, \apj, 928, 14, \dodoi{10.3847/1538-4357/ac51de}

\bibitem[{{Casas} {et~al.}(2022){Casas}, {Bonavera}, {Gonz{\'a}lez-Nuevo}, {Baccigalupi}, {Cueli}, {Crespo}, {Goitia}, {Santos}, {S{\'a}nchez}, \& {de Cos}}]{Casas22}
{Casas}, J.~M., {Bonavera}, L., {Gonz{\'a}lez-Nuevo}, J., {et~al.} 2022, \aap, 666, A89, \dodoi{10.1051/0004-6361/202243450}

\bibitem[{{Chandran} {et~al.}(2023){Chandran}, {Remazeilles}, \& {Barreiro}}]{Chandran23}
{Chandran}, J., {Remazeilles}, M., \& {Barreiro}, R.~B. 2023, arXiv e-prints, arXiv:2305.10193, \dodoi{10.48550/arXiv.2305.10193}

\bibitem[{{Delabrouille} \& {Cardoso}(2007)}]{DelabrouilleCardoso07}
{Delabrouille}, J., \& {Cardoso}, J.~F. 2007, arXiv e-prints, astro.
\newblock \doarXiv{astro-ph/0702198}

\bibitem[{{Delabrouille} {et~al.}(2009){Delabrouille}, {Cardoso}, {Le Jeune}, {Betoule}, {Fay}, \& {Guilloux}}]{Delabrouille09}
{Delabrouille}, J., {Cardoso}, J.~F., {Le Jeune}, M., {et~al.} 2009, \aap, 493, 835, \dodoi{10.1051/0004-6361:200810514}

\bibitem[{{Delabrouille} {et~al.}(2013){Delabrouille}, {Betoule}, {Melin}, {Miville-Desch{\^e}nes}, {Gonzalez-Nuevo}, {Le Jeune}, {Castex}, {de Zotti}, {Basak}, {Ashdown}, {Aumont}, {Baccigalupi}, {Banday}, {Bernard}, {Bouchet}, {Clements}, {da Silva}, {Dickinson}, {Dodu}, {Dolag}, {Elsner}, {Fauvet}, {Fa{\"y}}, {Giardino}, {Leach}, {Lesgourgues}, {Liguori}, {Mac{\'\i}as-P{\'e}rez}, {Massardi}, {Matarrese}, {Mazzotta}, {Montier}, {Mottet}, {Paladini}, {Partridge}, {Piffaretti}, {Prezeau}, {Prunet}, {Ricciardi}, {Roman}, {Schaefer}, \& {Toffolatti}}]{PlanckSkyModel}
{Delabrouille}, J., {Betoule}, M., {Melin}, J.~B., {et~al.} 2013, \aap, 553, A96, \dodoi{10.1051/0004-6361/201220019}

\bibitem[{{Eriksen} {et~al.}(2008){Eriksen}, {Jewell}, {Dickinson}, {Banday}, {G{\'o}rski}, \& {Lawrence}}]{Eriksen08}
{Eriksen}, H.~K., {Jewell}, J.~B., {Dickinson}, C., {et~al.} 2008, \apj, 676, 10, \dodoi{10.1086/525277}

\bibitem[{{Galloway} {et~al.}(2023){Galloway}, {Andersen}, {Aurlien}, {Banerji}, {Bersanelli}, {Bertocco}, {Brilenkov}, {Carbone}, {Colombo}, {Eriksen}, {Eskilt}, {Foss}, {Franceschet}, {Fuskeland}, {Galeotta}, {Gerakakis}, {Gjerl{\o}w}, {Hensley}, {Herman}, {Iacobellis}, {Ieronymaki}, {Ihle}, {Jewell}, {Karakci}, {Keih{\"a}nen}, {Keskitalo}, {Maggio}, {Maino}, {Maris}, {Mennella}, {Paradiso}, {Partridge}, {Reinecke}, {San}, {Suur-Uski}, {Svalheim}, {Tavagnacco}, {Thommesen}, {Watts}, {Wehus}, \& {Zacchei}}]{Galloway23}
{Galloway}, M., {Andersen}, K.~J., {Aurlien}, R., {et~al.} 2023, \aap, 675, A3, \dodoi{10.1051/0004-6361/202243137}

\bibitem[{{G{\'o}rski} {et~al.}(2005){G{\'o}rski}, {Hivon}, {Banday}, {Wandelt}, {Hansen}, {Reinecke}, \& {Bartelmann}}]{Healpix}
{G{\'o}rski}, K.~M., {Hivon}, E., {Banday}, A.~J., {et~al.} 2005, \apj, 622, 759, \dodoi{10.1086/427976}

\bibitem[{{Guilloux} {et~al.}(2007){Guilloux}, {Fay}, \& {Cardoso}}]{Guilloux07}
{Guilloux}, F., {Fay}, G., \& {Cardoso}, J.-F. 2007, arXiv e-prints, arXiv:0706.2598, \dodoi{10.48550/arXiv.0706.2598}

\bibitem[{{Han} {et~al.}(2021){Han}, {Sehgal}, \& {Villaescusa-Navarro}}]{Han21}
{Han}, D., {Sehgal}, N., \& {Villaescusa-Navarro}, F. 2021, \prd, 104, 123521, \dodoi{10.1103/PhysRevD.104.123521}

\bibitem[{{Kingma} \& {Ba}(2014)}]{Adam}
{Kingma}, D.~P., \& {Ba}, J. 2014, arXiv e-prints, arXiv:1412.6980, \dodoi{10.48550/arXiv.1412.6980}

\bibitem[{{Le Brun} {et~al.}(2015){Le Brun}, {McCarthy}, \& {Melin}}]{LeBrun15}
{Le Brun}, A.~M.~C., {McCarthy}, I.~G., \& {Melin}, J.-B. 2015, \mnras, 451, 3868, \dodoi{10.1093/mnras/stv1172}

\bibitem[{{Leach} {et~al.}(2008){Leach}, {Cardoso}, {Baccigalupi}, {Barreiro}, {Betoule}, {Bobin}, {Bonaldi}, {Delabrouille}, {de Zotti}, {Dickinson}, {Eriksen}, {Gonz{\'a}lez-Nuevo}, {Hansen}, {Herranz}, {Le Jeune}, {L{\'o}pez-Caniego}, {Mart{\'\i}nez-Gonz{\'a}lez}, {Massardi}, {Melin}, {Miville-Desch{\^e}nes}, {Patanchon}, {Prunet}, {Ricciardi}, {Salerno}, {Sanz}, {Starck}, {Stivoli}, {Stolyarov}, {Stompor}, \& {Vielva}}]{Leach08}
{Leach}, S.~M., {Cardoso}, J.~F., {Baccigalupi}, C., {et~al.} 2008, \aap, 491, 597, \dodoi{10.1051/0004-6361:200810116}

\bibitem[{{McCarthy} \& {Hill}(2023)}]{McCarthy23}
{McCarthy}, F., \& {Hill}, J.~C. 2023, arXiv e-prints, arXiv:2307.01043, \dodoi{10.48550/arXiv.2307.01043}

\bibitem[{{Piffaretti} {et~al.}(2011){Piffaretti}, {Arnaud}, {Pratt}, {Pointecouteau}, \& {Melin}}]{MCXC}
{Piffaretti}, R., {Arnaud}, M., {Pratt}, G.~W., {Pointecouteau}, E., \& {Melin}, J.~B. 2011, \aap, 534, A109, \dodoi{10.1051/0004-6361/201015377}

\bibitem[{{Planck Collaboration} {et~al.}(2016){Planck Collaboration}, {Aghanim}, {Arnaud}, {Ashdown}, {Aumont}, {Baccigalupi}, {Band ay}, {Barreiro}, {Bartlett}, {Bartolo}, {Battaner}, {Battye}, {Benabed}, {Beno{\^\i}t}, {Benoit-L{\'e}vy}, {Bernard}, {Bersanelli}, {Bielewicz}, {Bock}, {Bonaldi}, {Bonavera}, {Bond}, {Borrill}, {Bouchet}, {Burigana}, {Butler}, {Calabrese}, {Cardoso}, {Catalano}, {Challinor}, {Chiang}, {Christensen}, {Churazov}, {Clements}, {Colombo}, {Combet}, {Comis}, {Coulais}, {Crill}, {Curto}, {Cuttaia}, {Danese}, {Davies}, {Davis}, {de Bernardis}, {de Rosa}, {de Zotti}, {Delabrouille}, {D{\'e}sert}, {Dickinson}, {Diego}, {Dolag}, {Dole}, {Donzelli}, {Dor{\'e}}, {Douspis}, {Ducout}, {Dupac}, {Efstathiou}, {Elsner}, {En{\ss}lin}, {Eriksen}, {Fergusson}, {Finelli}, {Forni}, {Frailis}, {Fraisse}, {Franceschi}, {Frejsel}, {Galeotta}, {Galli}, {Ganga}, {G{\'e}nova-Santos}, {Giard}, {Gonz{\'a}lez-Nuevo}, {G{\'o}rski}, {Gregorio}, {Gruppuso}, {Gudmundsson}, {Hansen}, {Harrison},
  {Henrot-Versill{\'e}}, {Hern{\'a}ndez-Monteagudo}, {Herranz}, {Hildebrand t}, {Hivon}, {Holmes}, {Hornstrup}, {Huffenberger}, {Hurier}, {Jaffe}, {Jones}, {Juvela}, {Keih{\"a}nen}, {Keskitalo}, {Kneissl}, {Knoche}, {Kunz}, {Kurki-Suonio}, {Lacasa}, {Lagache}, {L{\"a}hteenm{\"a}ki}, {Lamarre}, {Lasenby}, {Lattanzi}, {Leonardi}, {Lesgourgues}, {Levrier}, {Liguori}, {Lilje}, {Linden-V{\o}rnle}, {L{\'o}pez-Caniego}, {Mac{\'\i}as-P{\'e}rez}, {Maffei}, {Maggio}, {Maino}, {Mandolesi}, {Mangilli}, {Maris}, {Martin}, {Mart{\'\i}nez-Gonz{\'a}lez}, {Masi}, {Matarrese}, {Melchiorri}, {Melin}, {Migliaccio}, {Miville-Desch{\^e}nes}, {Moneti}, {Montier}, {Morgante}, {Mortlock}, {Munshi}, {Murphy}, {Naselsky}, {Nati}, {Natoli}, {Noviello}, {Novikov}, {Novikov}, {Paci}, {Pagano}, {Pajot}, {Paoletti}, {Pasian}, {Patanchon}, {Perdereau}, {Perotto}, {Pettorino}, {Piacentini}, {Piat}, {Pierpaoli}, {Pietrobon}, {Plaszczynski}, {Pointecouteau}, {Polenta}, {Ponthieu}, {Pratt}, {Prunet}, {Puget}, {Rachen}, {Reinecke}, {Remazeilles},
  {Renault}, {Renzi}, {Ristorcelli}, {Rocha}, {Rossetti}, {Roudier}, {Rubi{\~n}o-Mart{\'\i}n}, {Rusholme}, {Sandri}, {Santos}, {Sauv{\'e}}, {Savelainen}, {Savini}, {Scott}, {Spencer}, {Stolyarov}, {Stompor}, {Sunyaev}, {Sutton}, {Suur-Uski}, {Sygnet}, {Tauber}, {Terenzi}, {Toffolatti}, {Tomasi}, {Tramonte}, {Tristram}, {Tucci}, {Tuovinen}, {Valenziano}, {Valiviita}, {Van Tent}, {Vielva}, {Villa}, {Wade}, {Wandelt}, {Wehus}, {Yvon}, {Zacchei}, \& {Zonca}}]{PlanckYMAP}
{Planck Collaboration}, {Aghanim}, N., {Arnaud}, M., {et~al.} 2016, \aap, 594, A22, \dodoi{10.1051/0004-6361/201525826}

\bibitem[{{Pratt} {et~al.}(2021){Pratt}, {Qu}, \& {Bregman}}]{Pratt21}
{Pratt}, C.~T., {Qu}, Z., \& {Bregman}, J.~N. 2021, \apj, 920, 104, \dodoi{10.3847/1538-4357/ac1796}

\bibitem[{{Remazeilles} {et~al.}(2013){Remazeilles}, {Aghanim}, \& {Douspis}}]{Remazeilles13}
{Remazeilles}, M., {Aghanim}, N., \& {Douspis}, M. 2013, \mnras, 430, 370, \dodoi{10.1093/mnras/sts636}

\bibitem[{{Sehgal} {et~al.}(2010){Sehgal}, {Bode}, {Das}, {Hernandez-Monteagudo}, {Huffenberger}, {Lin}, {Ostriker}, \& {Trac}}]{Sehgal10}
{Sehgal}, N., {Bode}, P., {Das}, S., {et~al.} 2010, \apj, 709, 920, \dodoi{10.1088/0004-637X/709/2/920}

\bibitem[{{Shallue} \& {Vanderburg}(2018)}]{Shallue18}
{Shallue}, C.~J., \& {Vanderburg}, A. 2018, \aj, 155, 94, \dodoi{10.3847/1538-3881/aa9e09}

\bibitem[{{Sunyaev} \& {Zeldovich}(1970)}]{SZ1970}
{Sunyaev}, R.~A., \& {Zeldovich}, Y.~B. 1970, \apss, 7, 3, \dodoi{10.1007/BF00653471}

\bibitem[{{Sunyaev} \& {Zeldovich}(1972)}]{SZ1972}
---. 1972, Comments on Astrophysics and Space Physics, 4, 173

\bibitem[{{Thorne} {et~al.}(2017){Thorne}, {Dunkley}, {Alonso}, \& {N{\ae}ss}}]{Thorne2017}
{Thorne}, B., {Dunkley}, J., {Alonso}, D., \& {N{\ae}ss}, S. 2017, \mnras, 469, 2821, \dodoi{10.1093/mnras/stx949}

\bibitem[{{Tr{\"o}ster} {et~al.}(2019){Tr{\"o}ster}, {Ferguson}, {Harnois-D{\'e}raps}, \& {McCarthy}}]{Troster19}
{Tr{\"o}ster}, T., {Ferguson}, C., {Harnois-D{\'e}raps}, J., \& {McCarthy}, I.~G. 2019, \mnras, 487, L24, \dodoi{10.1093/mnrasl/slz075}

\bibitem[{{Voskresenskaia} {et~al.}(2023){Voskresenskaia}, {Meshcheryakov}, \& {Lyskova}}]{Voskresenskaia23}
{Voskresenskaia}, S., {Meshcheryakov}, A., \& {Lyskova}, N. 2023, arXiv e-prints, arXiv:2309.17077.
\newblock \doarXiv{2309.17077}

\bibitem[{{Wagner-Carena} {et~al.}(2020){Wagner-Carena}, {Hopkins}, {Diaz Rivero}, \& {Dvorkin}}]{HGMCA}
{Wagner-Carena}, S., {Hopkins}, M., {Diaz Rivero}, A., \& {Dvorkin}, C. 2020, \mnras, 494, 1507, \dodoi{10.1093/mnras/staa744}

\bibitem[{{Zhu} {et~al.}(2019){Zhu}, {Dai}, {Bian}, {Chen}, {Chen}, \& {Hu}}]{Zhu19}
{Zhu}, X.-P., {Dai}, J.-M., {Bian}, C.-J., {et~al.} 2019, \apss, 364, 55, \dodoi{10.1007/s10509-019-3540-1}

\end{thebibliography}



\end{document}